\documentclass[aps,twocolumn,floats,pre]{revtex4-1}
\usepackage{graphics,graphicx,epsfig}
\pdfoutput=1
\usepackage{amssymb,color}
\usepackage{epsf,epstopdf,wrapfig}
\usepackage{amsmath}

\begin{document}

\title{Emergence of collective changes in travel direction of starling flocks\\
from individual birds fluctuations}

\author{Alessandro Attanasi$^{1,2}$}
\author{ Andrea Cavagna$^{1,2,3}$} 
\author{Lorenzo Del Castello$^{1,2}$} 
\author{Irene Giardina$^{1,2,3\, \ddagger}$}
\author{Asja Jelic$^{1,2\, \ddagger}$}
\author{Stefania Melillo$^{1,2}$}
\author{Leonardo Parisi$^{1,2,4}$}
\author{Oliver Pohl$^{1,2}$}
\email{Present address: Institut f\"ur Theoretische Physik, Technische Universit\"at Berlin, Hardenbergstrasse 36, D-10623 Berlin-Charlottenburg, Germany.}
\author{Edward Shen$^{1,2}$}
\author{Massimiliano Viale$^{1,2}$}

\affiliation{$^1$ Istituto Sistemi Complessi, Consiglio Nazionale delle Ricerche, UOS Sapienza, 00185 Rome, Italy}

\affiliation{$^2$ Dipartimento di Fisica, Universit\`a\ Sapienza, 00185 Rome, Italy}

\affiliation{$^3$ Initiative for the Theoretical Sciences, The Graduate Center, 
City University of New York, 10016 New York, USA}

\affiliation{$^4$ Dipartimento di Informatica, Universit\`a\ Sapienza, 00198 Rome, Italy}

\affiliation{$^\ddagger$ e-mail: asja.jelic@gmail.com, irene.giardina@roma1.infn.it}

\begin{abstract}
One of the most impressive features of moving animal groups  is their ability to perform sudden coherent changes in travel direction. While this collective decision can be
a response to an external perturbation, such as the presence of a predator, recent studies show that such directional switching can also emerge from the intrinsic fluctuations in the individual behaviour. 
However, the cause and the mechanism by which such collective changes of direction occur are not fully understood yet. 
Here, we present an experimental study of spontaneous collective turns in natural flocks of starlings. We employ a recently developed tracking algorithm to reconstruct three-dimensional trajectories of each  individual bird in the flock for the whole duration of a turning event. Our approach enables us to analyze changes in the individual behavior of every group member and reveal the emergent dynamics of turning. We show that spontaneous turns start from individuals located at the elongated edges of the flocks, and then propagate through the group. 
We find that birds on the edges deviate from the mean direction of motion much more frequently than other individuals, indicating that persistent localized fluctuations are the crucial ingredient for triggering a collective directional change.  Finally, we quantitatively show that birds follow equal radius paths during turning allowing the flock to change orientation and redistribute risky locations among group members.  The whole process of turning is a remarkable example of how a self-organized system can sustain collective changes and reorganize, while retaining coherence. 
\end{abstract}

\maketitle


\section{Introduction}

Moving animal groups are a paradigmatic example of collective behavior in social species.
The most striking features of such a collective motion are rapid, coherent changes in the 
direction of travel of the whole group. 
Each such change involves a collective decision that starts with a consensus-forming  mechanism leading to the decision, and its actual execution by propagation through the  entire group. 
At any moment of this process, the stakes of decreasing the fitness of any individual of the group are high.
Both, slightest uncertainty and a slow and inefficient transport of information, are punished 
by decrease of cohesion, or even splitting of the group, leaving some individuals an easy 
pray to the predator. 
Determining the factors that govern the collective change of direction, and the mechanism 
ensuring efficient propagation of this collective decision, is thus a key to understanding the 
animal movement in groups. 

Sudden collective changes of state in animal groups happen often \cite{couzin+krause_03,sumpter_06,buhl_06,radakov_1973,tunstrom_13}. Sometimes they may be a result of a global alarm cue, such is a shot heard by an entire flock of birds. 
In this case, the collective change of state is not necessarily a social response of the group. There is basically no transfer of information between the group members -- all the birds react at the same moment, apart from individual differences in response times.
More fascinating are collective decisions that have a localized spatial origin, starting from a few individuals that are close to each other.  Consensus is achieved between nearby group members and a decision is formed, then the information to  change state travels across the whole group and reaches all individuals.

In some cases, the cause of the localized spatial origin may be an external stimulus, such as a nearby predator that is seen by a small number of individuals in the group. 
Nevertheless, it has been shown that the collective directional switching can be also triggered 
spontaneously, without changes in the external environment \cite{buhl_06}. 
In fact, observation of natural starling flocks reveals that most of the time the flocks turn in absence of predators and with no apparent reason.
Why, then, a collective turn occurs in the first place? What causes the first nearby birds to initiate a turn, and the others to follow? What are the consequences of turning?

Here we perform an experimental study in which we address these intriguing questions about  collective turns in natural flocks of starlings. Using detailed data on individual trajectories in large groups we reveal the mechanism responsible for the initiation of spontaneous turning in starling flocks. 
We find that such turns always start from the elongated edges of the flock. This result suggests that the turn occurs because individual birds which are positioned in specific locations -- with higher risks for predatory attacks and lesser social feedback from neighbors -- are more prone to rearrange their position. In fact we find that the birds which initiate the turn display unusual deviations from the mean flock's motion over longer periods than other birds. This persistent signal provokes a response of the neighboring birds, which leads to a decision to turn.
Finally, we characterize the kinematics of turning and show that individuals follow equal radius paths. As a consequence, birds change their position with respect to the global direction of motion: dangerous edge locations become front/back ones and risk is redistributed through the group.

\section{Results }

\begin{figure}[t]
  \centering
  \includegraphics[width=1\columnwidth,]{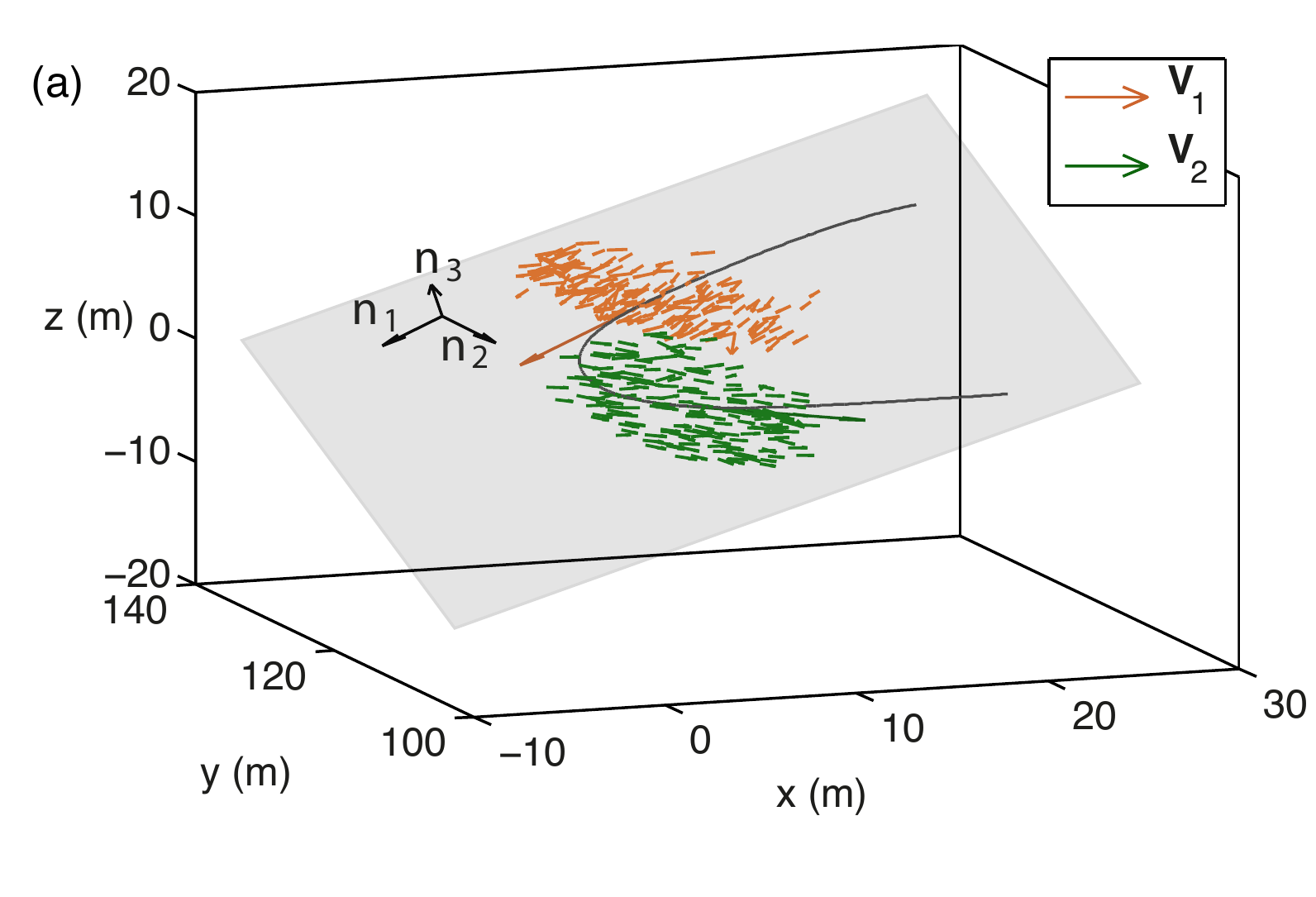}\\[0.1cm]
   \includegraphics[width=1\columnwidth,]{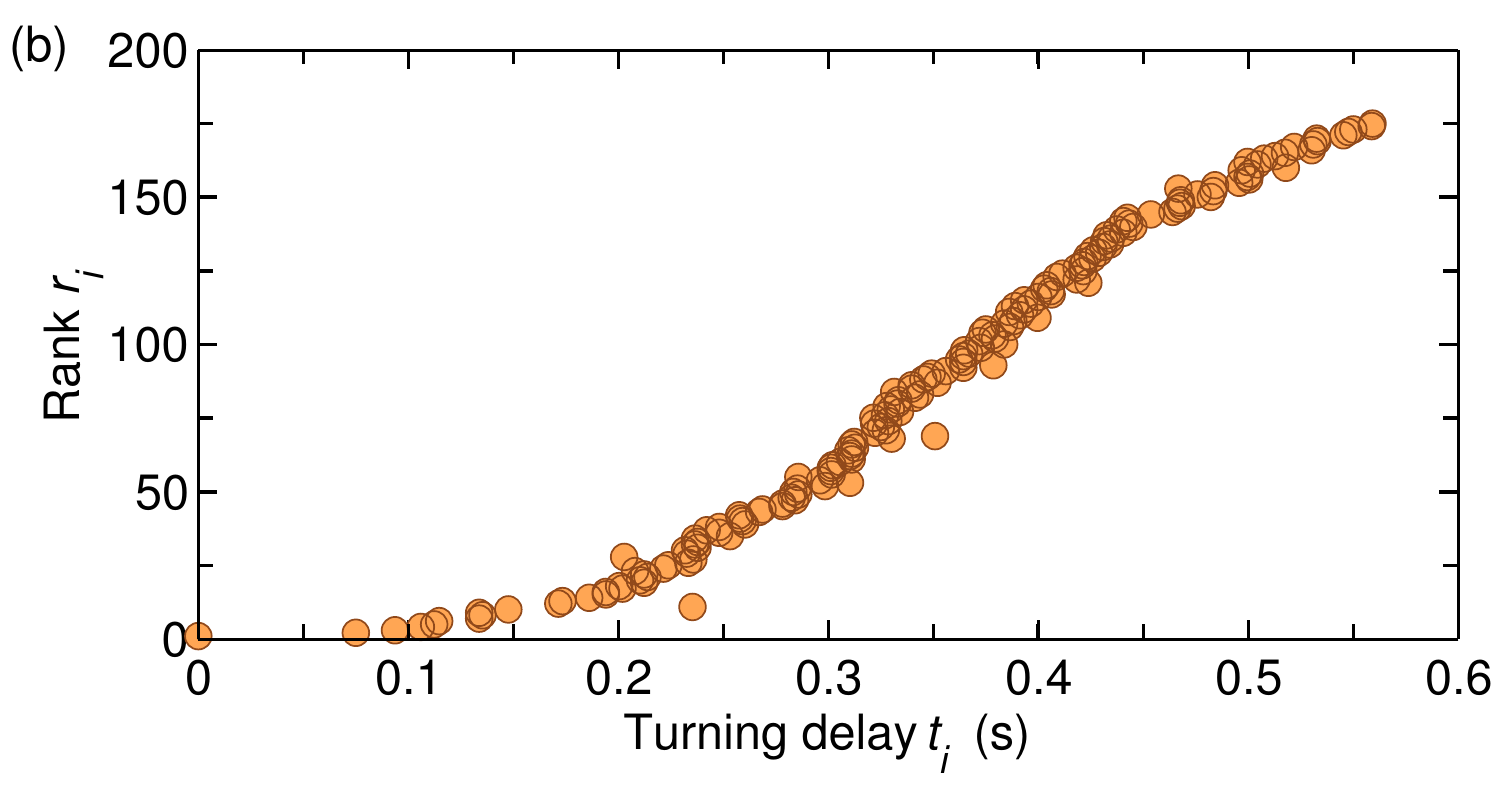}
 \caption{{\bf Collective turn and ranking.} 
(a) For a flock performing a collective turn (event E1 in Table \ref{table:flocks}), a trajectory of its barycenter is shown (black curve). 
Individual birds and their velocity vectors are shown at moment $t_1$ of the start of the turn (orange velocity vectors) and moment $t_2$ after the turn is finished (green velocity vectors). Before and after the turn is performed, the flock's trajectory is almost straight. 
Average flock's velocities ${\bf V}_1$ and ${\bf V}_2$ at times $t_1$ and $t_2$, respectively, are shown. Note that these velocity vectors are rescaled (increased), in order to emphasize the flock's direction of motion before and after the turn.
The flock's `turning plane' is shown in gray and determined by an orthogonal coordinate system  $({\bf n}_1,{\bf n}_2,{\bf n}_3)$ defined in the main text and SI.
(b) The rank $r_i$ of each bird $i$, i.e.\ its order in the turning sequence, is plotted vs its turning time delay $t_i$, i.e. the delay with respect to the first bird to turn (ranking curves for all turning events are given in ref.~\cite{attanasi+al_14}).
 }
\label{fig:turning-plane}
\end{figure}

While returning to their roost shortly before sunset, starlings form sharp-bordered flocks which wheel and turn over the roosting place before setting down to the trees. They perform highly synchronized maneuvers while maintaining strong coherence, either as  a response to a predator attack, but most often with no visible external influence.
We concentrate exactly on the latter type of collective turns, where no changes in external environment are observed (Video 1). 
We collected data for $12$ turning events and tracked   in time the positions of all individual birds in a flock.
The full 3D dynamical trajectory of each bird in the flock is obtained during the whole duration of the turn by using a 3-cameras setup and a tracking algorithm developed in ref.~\cite{attanasi+al_13b} (see Methods for more information). 
In Fig.\ref{fig:turning-plane}-a and in Video 1 we show a typical collective turn. 

Having the full reconstructed trajectories, we can rank all birds in the flock according to their turning order, using the ranking procedure developed in ref.~\cite{attanasi+al_14}. 
We can therefore say who is the first to turn, who is second, and so on. In this ranking, each bird $i$ is labelled by its rank $r_i$, and by its absolute turning time $t_i$, that is the turning delay with respect to the top bird in the rank -- the initiator. 
The ranking curve $r(t)$ in Fig.\ref{fig:turning-plane}-b is obtained by plotting the rank $r_i$ of each bird as a function of its absolute turning time $t_i$.

%
%

In \cite{attanasi+al_14} we showed that the ranking curves $r(t)$ for all turning events summarized in Table \ref{table:flocks} take similar shape, indicating that turns are initiated by a small number of birds, whose reaction times are relatively long as the turn starts.  Moreover, the first birds to turn are physically close to each other.  Hence, the decision to turn has a spatially localized origin and it then propagates across the flock through a {\it social} transfer of information from bird to bird.  This information flow is illustrated in Video 2 in which each bird changes color (from grey to red) once it starts turning, resulting in a turning wave that propagates through the whole flock. 


The mechanism through which such propagation occurs is by far non trivial, with directional information traveling undamped with a speed of propagation that is larger the more ordered the flock is. In \cite{attanasi+al_14} we quantified in detail the features of this propagation, which is not described by standard models of self-propelled motion, and introduced a new mathematical theory able to explain it \cite{attanasi+al_14,cavagna+al_14}.  
There are, however, a few fundamental questions that remain to be addressed:  {\it why} such a collective change of state  started in the first place; who and why initiated the turn; and what are the consequences of turning in terms of global structure and individual rearrangements.


\begin{figure*}[t!]
  \centering
   \includegraphics[width=1.9\columnwidth,]{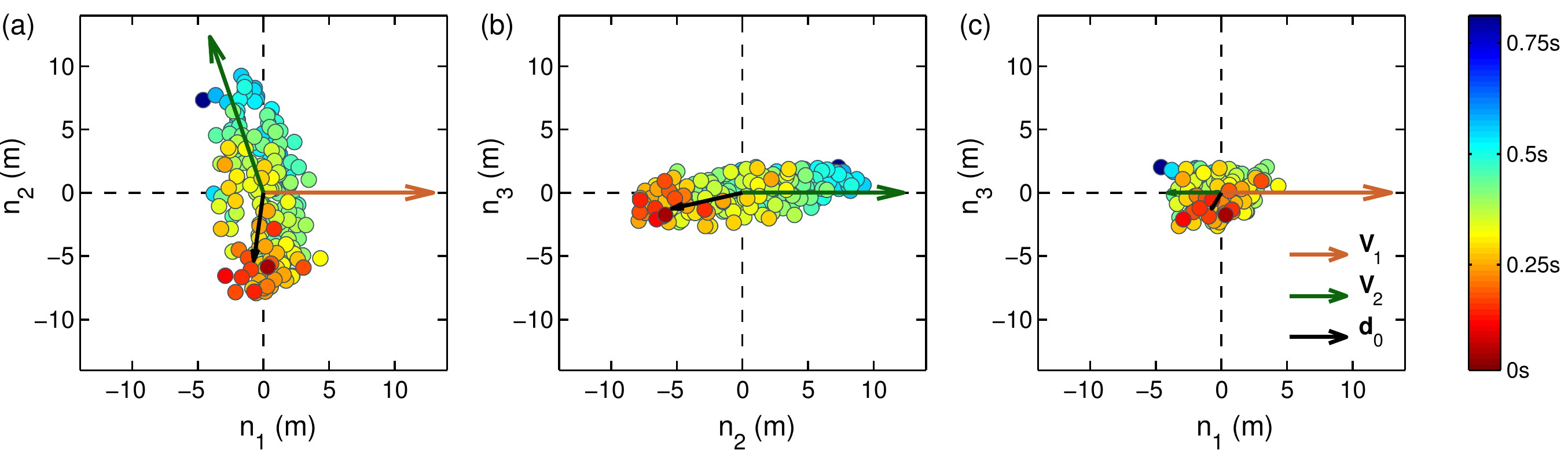}
 \caption{{\bf Propagation of the turn.} 
Flock 
E1 with 176 birds is shown at moment  $t_1$ of the start of the turn 
by using its projections on the planes of the `turning plane' orthogonal system $({\bf n}_1,{\bf n}_2,{\bf n}_3)$ centered at the flock's center of mass. 
Different panels show a view on the flock (a) from the top, (b) from the front, and (c) from a side.
At time $t_1$ the flock is moving along the direction ${\bf n}_1$ with velocity ${\bf V}_1$, which is to the right in panels (a) and (c), while in panel (b) it moves towards the viewer. At moment $t_2$ after the turn, the flock is moving with the velocity ${\bf V}_2$. The velocity vectors are rescaled (increased), in order to emphasize the old and new direction of global flock's motion. 
A wave of directional change information spreads through the flock during a turn. The birds are colored according to their turning time delays $t_i$, as indicated by the colorbar on the right. The mean position of the first $10$ birds to turn at time $t_1$, with respect to the barycenter, is indicated by vector ${\bf d}_0$.
Propagation of the directional information for several other turning flocks is shown in Supplementary Information.
 }
\label{fig:turn-propagation-rainbow}
\end{figure*}

To investigate these questions, we first look at the spatial position of the top ranked birds within the flock and how from there the directional information propagates through the group.
Flocks have three-dimensional shapes and it is therefore difficult to represent the spatial modulation of the turn propagation by using two-dimensional figures.  
To simplify the analysis, we exploit the fact that the trajectories of birds during a turn lie approximately on a plane (see Fig.\ref{fig:turning-plane}-a and Fig.\ref{fig:velocity-scalar-prod}). We can conveniently define a coordinate system with unit vectors $({\bf n}_1,{\bf n}_2,{\bf n}_3)$, where $({\bf n}_1,{\bf n}_2)$ lie on the turning plane and ${\bf n}_3$ is perpendicular to it, with ${\bf n}_1$ being the direction of motion of the flock at the beginning of the turn (see SI for details). This coordinate system is very useful to visualize turns. An example is given in Fig.\ref{fig:turn-propagation-rainbow}, where we show the birds positions at moment $t_1$ of the start of the turn.

At time $t_1$ the flock is moving with the velocity ${\bf V}_1$ (along the ${\bf n}_1$ axis).
Velocity ${\bf V}_2$ indicates a new flight direction, which is assumed after the collective turn is performed.
For a flock of $N$ birds, flock's velocity is defined as  ${\bf V}(t)=(1/N) \sum_i {\bf v}_i(t)$, where ${\bf v}_i(t)$ is velocity of bird $i$ at time $t$.
The birds are colored according to their turning time delays, revealing a response chain from the first to the last bird. This figure shows a few very interesting facts.

The top ranked birds (colored in red) are close to each other and  are located close to one of the elongated edges of the flock. Once the turn starts, the information propagates from the initiating birds in all directions, ending with the birds close to the opposite elongated edge of the flock.  As in Video 2, spatial modulation of the turning wave indicates a social nature of the response.
Finally, the orientation of the flock with respect to the direction of motion changes upon turning, with the flock initially moving perpendicularly to its longest axis and ending with a direction of motion parallel to it.  All the features we have qualitatively described so far can be precisely quantified. As we shall discuss in the remaining of the paper they hold systematically in all the turning events we have analyzed, indicating that spontaneous turns occur with a generic mechanism.

\subsection{Origin of the turn}

\begin{figure}[t!] 
  \centering
     \includegraphics[width=1 \columnwidth]{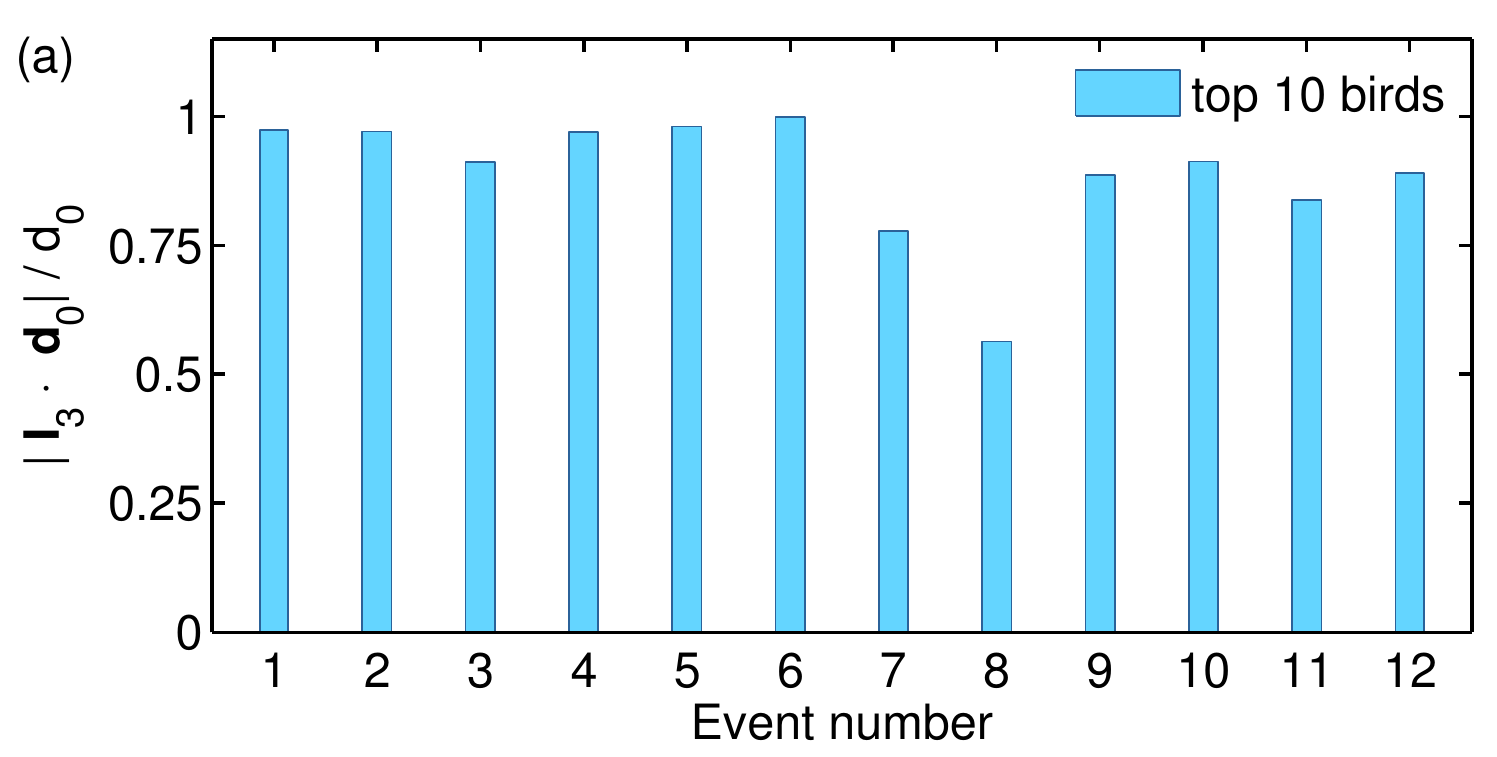} \\
     \includegraphics[width=1 \columnwidth]{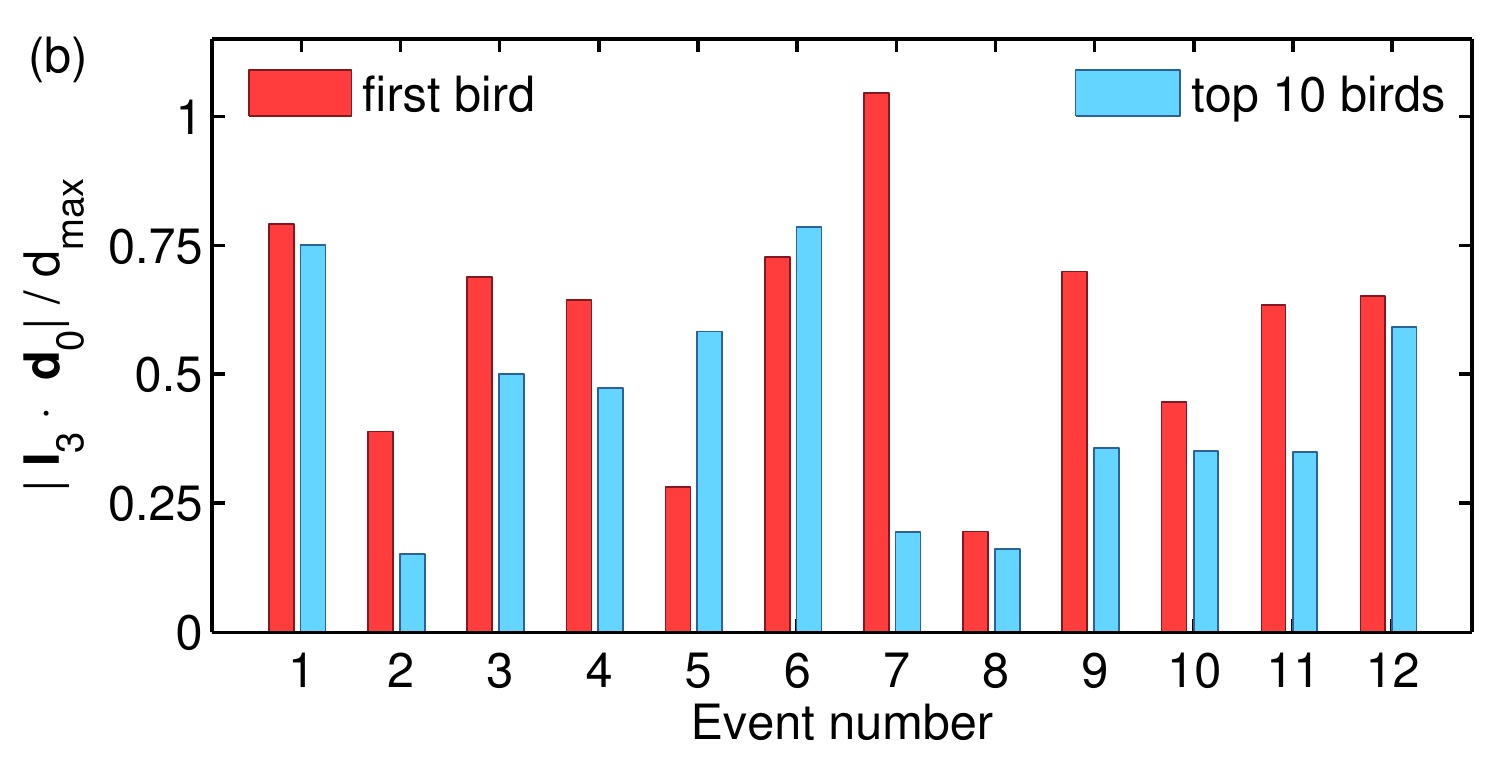}  
 \caption{
{\bf Position of the origin of the turn.}
We examine the position of the top-ranked birds that initiate the turn with respect to the longest elongation axis ${\bf I}_3$ at the start of the turn $t_1$ for each turning event.
(a) Absolute scalar product of the unitary vector ${\bf I}_3$ with the unitary vector of the mean position relative to the barycenter 
${\bf d}_0/d_0$ of the $10$ top-ranked birds (data are given in Supplementary Table S1).  
The result does not change qualitatively for another number of top-ranked birds, as long as this number is small compared to the total number of birds in the flock $N$. 
Large values of this scalar product confirm that the origin of the turn is positioned along the longest elongation axis.
(b) Scalar product $\left|{\bf I}_3\cdot{\bf d}_0\right|/d_{\rm max}$, where $d_{\rm max}$ is the maximal possible distance of the birds along the ${\bf I}_3$ axis at the edge where the turn started. We show the results for ${\bf d}_0$ calculated as a mean position of the $10$ top-ranked birds (blue bars), but also as a position of only the first bird that started the turn (red bars). For most events, the values are higher than $0.5$, although not too close to $1$ due to the sparseness of the flock at the very edges and particular orientations. Together with Fig.\ref{fig:turn-propagation-rainbow} and Fig.\ref{fig:turn-propagation-rainbow-3flocks}, this shows that the origin of the turn is close to the elongated edges.
}
\label{fig:nucleus}
\end{figure}

We can quantitatively locate the origin of the turn by looking at the positions of the first birds that turn within the flock. 
Let us define the average position of the $10$ top-ranked birds at the start of the turn $t_1$ with respect to the barycenter of the flock:
\begin{equation}
{\bf d}_0 = \frac{1}{10}\sum_{i=1}^{10}\left ({\bf r}_i - {\bf r}_{\rm cm}\right)
\label{eq:d0}
\end{equation}
where ${\bf r}_i$ is position of bird with rank $i$ at $t_1$, and ${\bf r}_{\rm cm} = (1/N) \sum_i {\bf r}_i $ is position of the center of mass of the flock. 
In Fig.\ref{fig:turn-propagation-rainbow} we can see that the first turning birds are positioned along the longest axis of the flock and close to its edge. To  quantify this behavior, we compute the  absolute scalar product of the normalized vector ${\bf d}_0/d_0$ with the unitary vector ${\bf I}_3$ of the longest elongation axis at the start of the turn $t_1$
(see SI for details on flocks' elongation axes).  This quantity gives a measure of the orientation of the initiating birds with respect to ${\bf I}_3$ and is displayed for all the turning events in Fig.\ref{fig:nucleus}-a  (see also Table \ref{table:flocks}).
The large values (close to $1$) indicate that the first birds to turn are situated along the longest elongation axis ${\bf I}_3$ for all the analyzed turns.
To quantify how close these birds are to the lateral tips of the flock, we calculate $\left|{\bf I}_3\cdot{\bf d}_0\right|/d_{\rm max}$, where $d_{\rm max} = \max \{({\bf r}_i-{\bf r}_{\rm cm})\cdot {\bf I}_3\}$ is the maximal possible value of a bird's distance along ${\bf I}_3$ axis, with $i$ going over all the birds on the side of the flock at which the turn started (since the two edges could be of different lengths). In Fig.\ref{fig:nucleus}-b, we show the values of this quantity for both the first bird to start the turn and for the 10 top-ranked birds. The obtained values are quite high, confirming that initiators are typically close to the lateral elongated edges. 

In Table \ref{table:flocks}, we give values for other quantities that give additional information about the position of the origin of the turn. 
For example, we can look at the normalized scalar product between ${\bf d}_0$ and the flock's velocity ${\bf V}_1$ at the start of the turn (see column ${\bf V}_1\cdot {\bf d}_0$ in Table \ref{table:flocks}). Its low values indicate that in most events the initiating birds are situated on the sides of the flock  and not in the front or in the back, consistent with our previous analysis of Fig.\ref{fig:nucleus}.
In the same way, we can calculate the normalized scalar product between ${\bf d}_0$ and the velocity ${\bf V}_2$ assumed by the flock {\it after} the turn  (column ${\bf V}_2\cdot {\bf d}_0$ in Table \ref{table:flocks}). For most of the flocks, this product is negative suggesting that once the top ranked birds initiate the turn they try to move towards the center of the flock, therefore ``pushing'' the whole flock to turn in the direction opposite to vector ${\bf d}_0$ as visually displayed in the flocks' projections of Fig.\ref{fig:turn-propagation-rainbow} and Fig.\ref{fig:turn-propagation-rainbow-3flocks}.


\begin{figure}[t!] 
  \centering
     \includegraphics[width=0.5 \columnwidth]{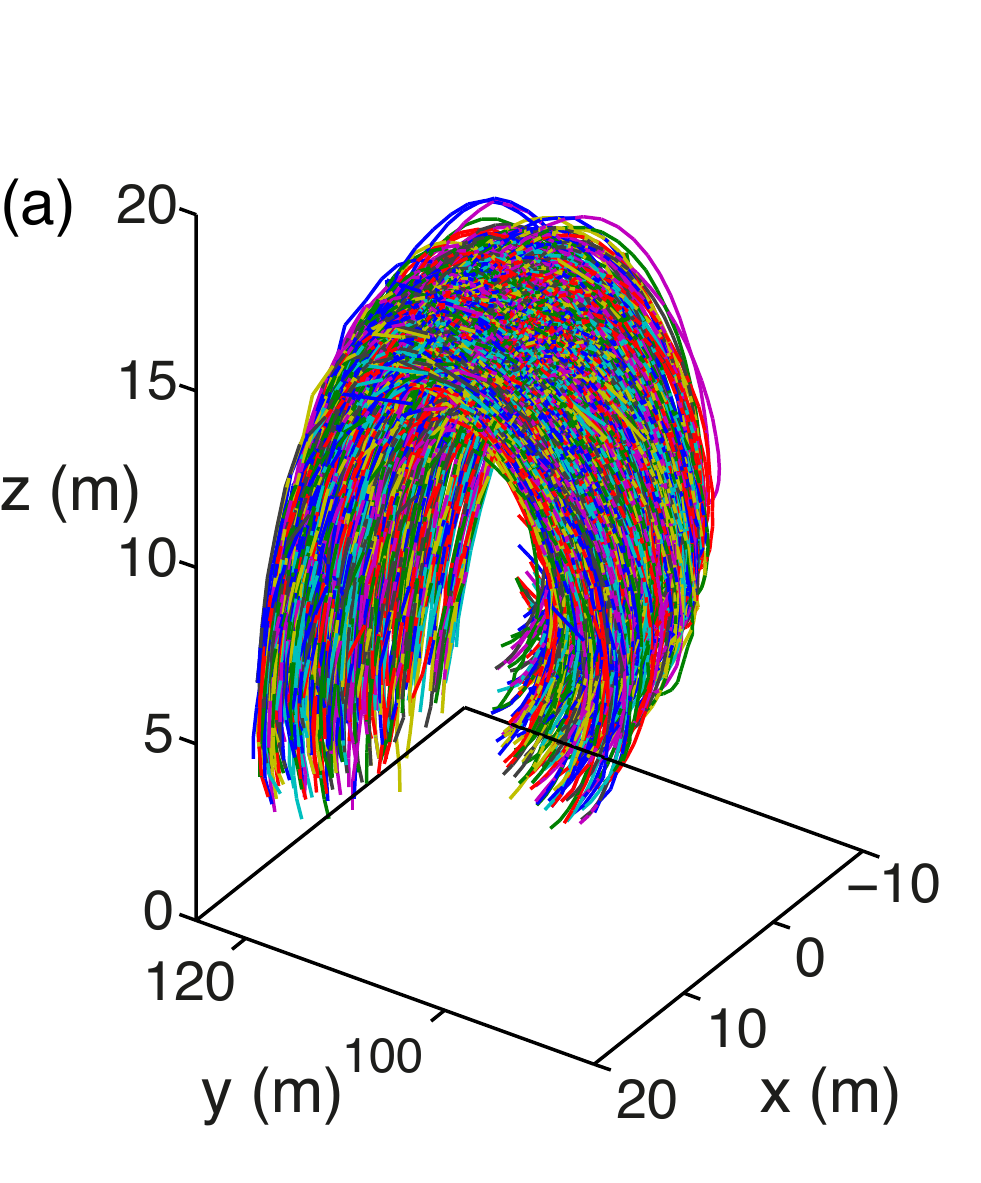} 
    \includegraphics[width=0.4\columnwidth]{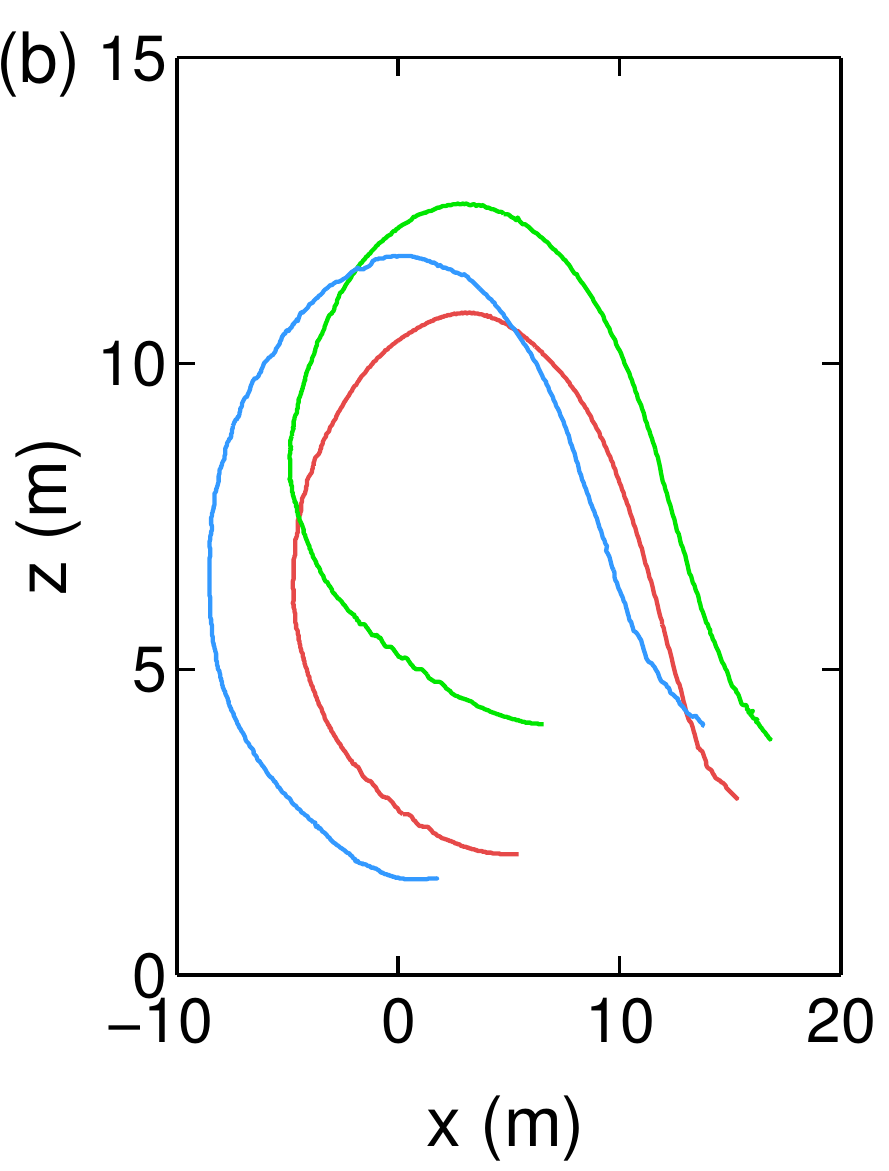}
 \caption{
{\bf Equal radius trajectories.}
(a) Three dimensional reconstruction of the full individual trajectories during the turning event E6 in Table \ref{table:flocks}.
(b) Three individual trajectories of nearby birds: the equal radius paths are clearly visible, each bird turning around a different rotation center but with the same radius of curvature of its neighbors.}
\label{fig:trajectories}
\end{figure}

\subsection{Equal radius paths and global reorientation during turns}

Once the turn initiates at the edges of the flock, a turning wave propagates through the group according to the ranking curve displayed in Fig.\ref{fig:turning-plane}-b.  During this process, two things happen: 
each bird performs its own individual turn following a specific trajectory in space; and the flock as a whole performs a global collective turn. These two dynamics are in fact strictly interconnected, and the way individuals coordinate turning with each other determines how the flock turns as a whole. Let us now show in detail how this occurs.

A few experimental and numerical studies in the past \cite{ballerini+al_08b,heppner_92,cavagna+al_14,hemelrijk+al_10} observed that, when turning, birds follow equal  radius paths. Having the full 3D trajectories of all individuals, we can investigate in detail this issue. In Fig.\ref{fig:trajectories}  we report the individual trajectories during a turn, which clearly show very similar radii of curvature and trajectory crossing.  A quantitative investigation of how the whole network changes during the turn confirms that birds indeed follow equal radius turning (see SI and Fig.\ref{fig:equal-radius}).
Our analysis in \cite{attanasi+al_14,cavagna+al_14} also shows that this kind of turning is intimately related to the fast and efficient way turns occur, where each bird starting to turn transfers to its neighbors information on its direction of motion and path curvature  through a social interaction mechanism, giving rise to a propagating wave of turning individuals. 

Equal radius turning is thus very non-trivial. It is also completely different from how a rigid assembly of particles (like a plane, or any solid body) would turn: in that case all particles turn synchronously around the same rotation point, following parallel paths and having different speeds and radii of curvature. Birds in a flock, on the contrary, turn following a social transfer of information, using different rotation points, but with the same radius of curvature and speed.  Interestingly, this way of turning is advantageous under many respects. 
On the one hand, individuals can  keep approximately constant speed and  produce amazingly quick collective turns. On the other hand, as we shall discuss, the reciprocal positions of individuals and the orientation of the flock in space change so as to redistribute boundary locations  and alternate risk.

\begin{figure}[t!] 
  \centering
    \includegraphics[width=1 \columnwidth]{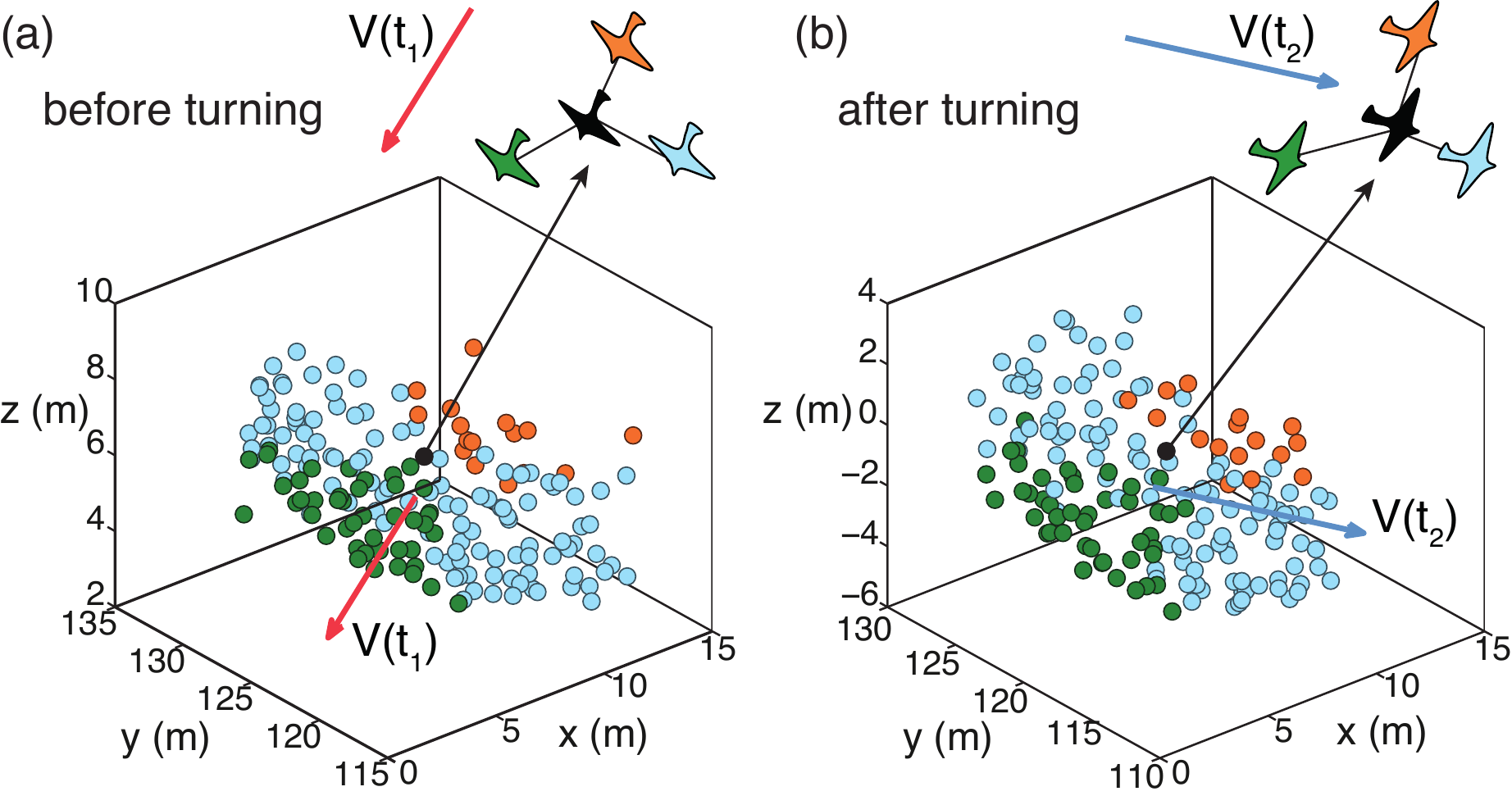}\\ [0.2cm]
    \includegraphics[width=1 \columnwidth]{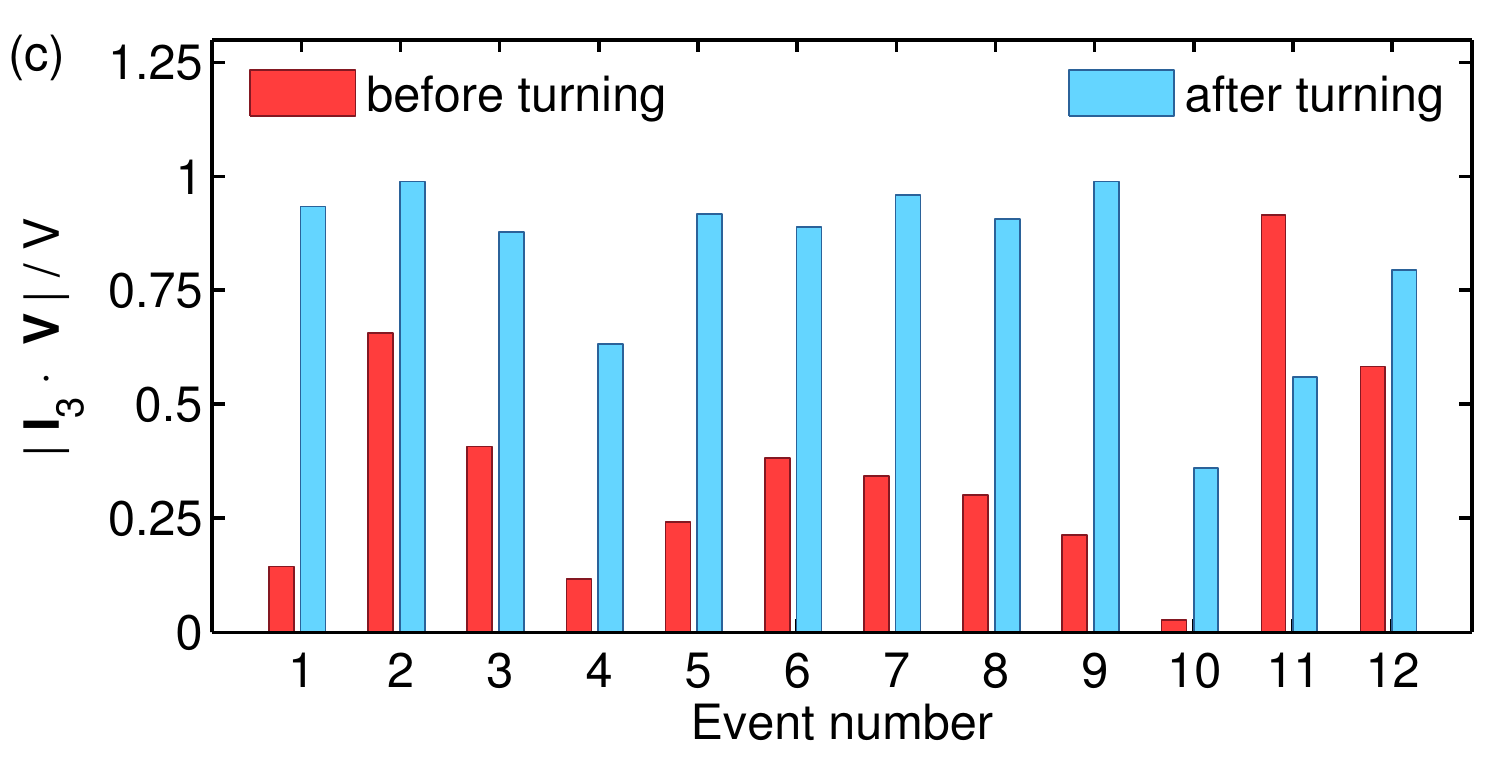}
 \caption{
{\bf Reorientation of the flock during the turn.}
(a) and (b): Change of the orientational topology around a reference bird in the flock. Panel (a) shows the flock at the start of the turn. The reference bird is pictured in black, birds flying in front with respect to the bird are green those on the sides are in light blue, the ones behind are in orange. Panel (b) shows the flock at the end of the turn: birds are pictured with the same colors as in panel (a).
(c) Global orientation of the flock before and after performing a turn. We report scalar products between 
unitary vector ${\bf I}_3$ of the flock's longest elongation axis with its direction of motion, i.e.\ more precisely with a unit vector of velocity ${\bf n}(t) = {\bf V}(t)/V(t)$, at times $t_1$ (start of the turn)  and $t_2$ (end of the turn).
Data are given in Supplementary Table S1.
The angle between the longest elongation axis and the direction of motion is large before the turn and it reduces during the turn. Moreover, for almost all events, the initial values of this scalar product are below $0.5$, i.e.\ the angle between the longest elongation axis and flight direction is $60^{\rm o}$ or larger.
The exceptions are events E11 and E12: these events correspond to two consecutive turns of the same flock, and occur after a merging of two separate flocks into one (see SI), which might be the reason  why the reorientation is different. 
}
\label{fig:elongation}
\end{figure}

To understand how this occurs, let us look at Fig.\ref{fig:elongation}. In panel (a), we show a reconstructed flock at the initial time $t_1$ of the turn. 
A random bird is plotted in black, and the other birds are colored according to their angular position around this reference bird with respect to the barycenter velocity: green for the birds flying in front, light-blue for the ones on the sides, and orange for the birds flying behind. In panel (b) we show the same flock at time $t_2$ after performing a turn of $120\mathrm{^{\circ}}$, with the same colors previously assigned. The two figures reveal that the orientational topology changes with the change of the direction of motion, e.g., the birds which were in front of the reference bird (green) are now flying at the right of it. At the same time, the structural network of birds remains stable, that is, their angular relative position does not change considerably (see SI). 

When looking at the flock as a whole, this implies that the group retains the same orientation with respect to an absolute reference frame, but its overall orientation with respect to the direction of motion changes.  As one can see in Fig.\ref{fig:turn-propagation-rainbow} (see also Fig.\ref{fig:turn-propagation-rainbow-3flocks}), shortly before the turn the flock has its longest elongation axis perpendicular to the direction of motion ${\bf V}_1$ at time $t_1$. After the turn at time $t_2$, however, its new direction of motion ${\bf V}_2$ (green arrow in Fig.\ref{fig:turn-propagation-rainbow})  is not anymore orthogonal to the longest elongation axis, but rather parallel to it. This behavior is characteristic for almost all the analyzed flocks . To quantify it, we computed  the normalized scalar product of the longest elongation axis ${\bf I}_3$ and the flock's velocity before and after the turn, i.e.\ at times $t_1$ and $t_2$, for all our turning events. The result is displayed in Fig.\ref{fig:elongation}-c  and shows that  flocks  tend to fly with their longest axis perpendicular to the direction of motion prior to turns (small ${\bf I}_3\cdot {\bf V}_1$), while align this axis with the flock's velocity after the turn  (large ${\bf I}_3\cdot {\bf V}_2$). 
A full analysis of the dynamical evolution of the three inertial axis during turns confirms this scenario (see SI and Fig.\ref{fig:velocity-scalar-prod}).

Interestingly, the flock's reorientation during the turn significantly changes the positional role of individuals in the group. Birds that were located at the largely populated front or back boundaries of the flock end up on the side, while birds on the lateral sparse tips, i.e. the same birds that initiated the turn in the first place, move to the front or back.

\subsection{Individual deviations from the global direction}\label{section:deviations}


In the previous sections we have characterized the kinematics of turning. We have shown where turns originate in the flock and how, due to the specific equal radius way of turning, individuals rearrange their positions with respect to direction of motion. What we would like to investigate in the remaining sections is the mechanism that triggers the start of the turn. 

As we have seen, turns initiate spontaneously from the lateral elongated edges of the flock. There might be some important difference in the way birds located in these regions of the flock behave that can explain why this occurs. To elucidate this point we look in more detail at the characteristics of individual motion, and its variability through the flock. More specifically, since turns involve a permanent change in the direction of motion, we can quantify the individual tendency to deviate from the global flock's direction prior to the occurrence of the  turn itself.

\begin{figure}[t!] 
  \centering
\includegraphics[width=.95 \columnwidth]{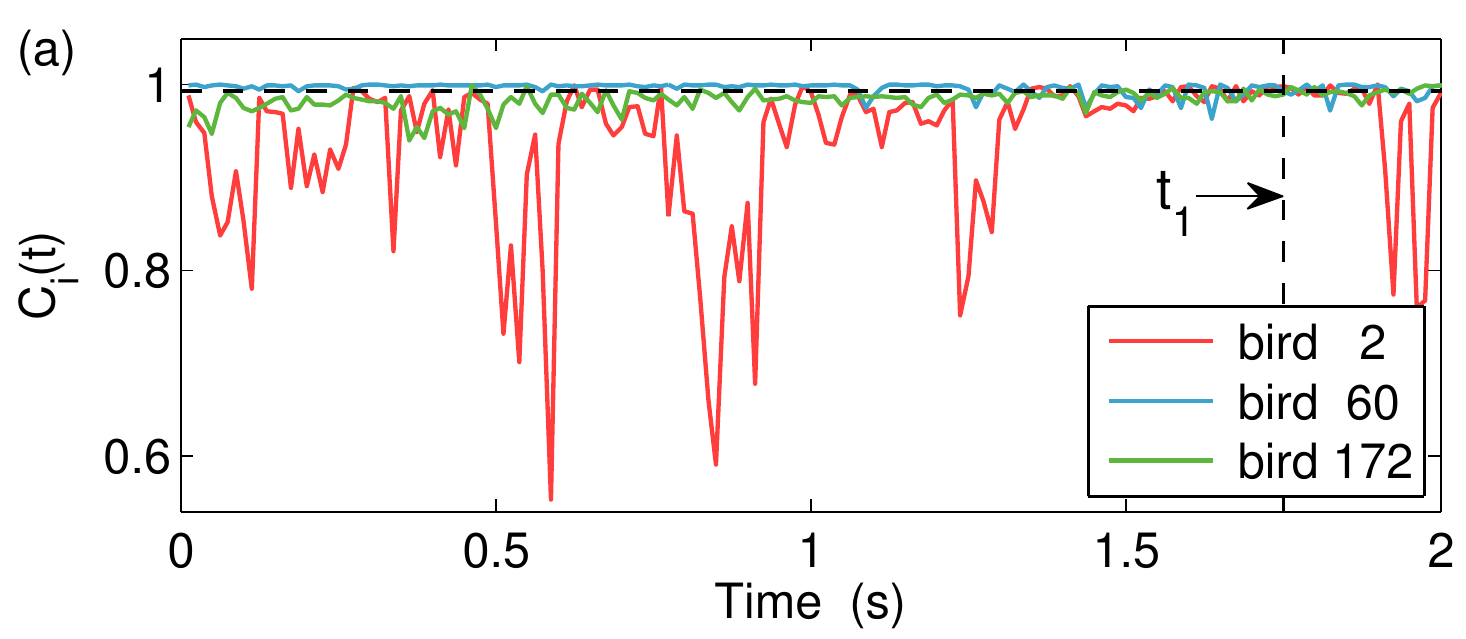}\\  
\includegraphics[width=.95 \columnwidth]{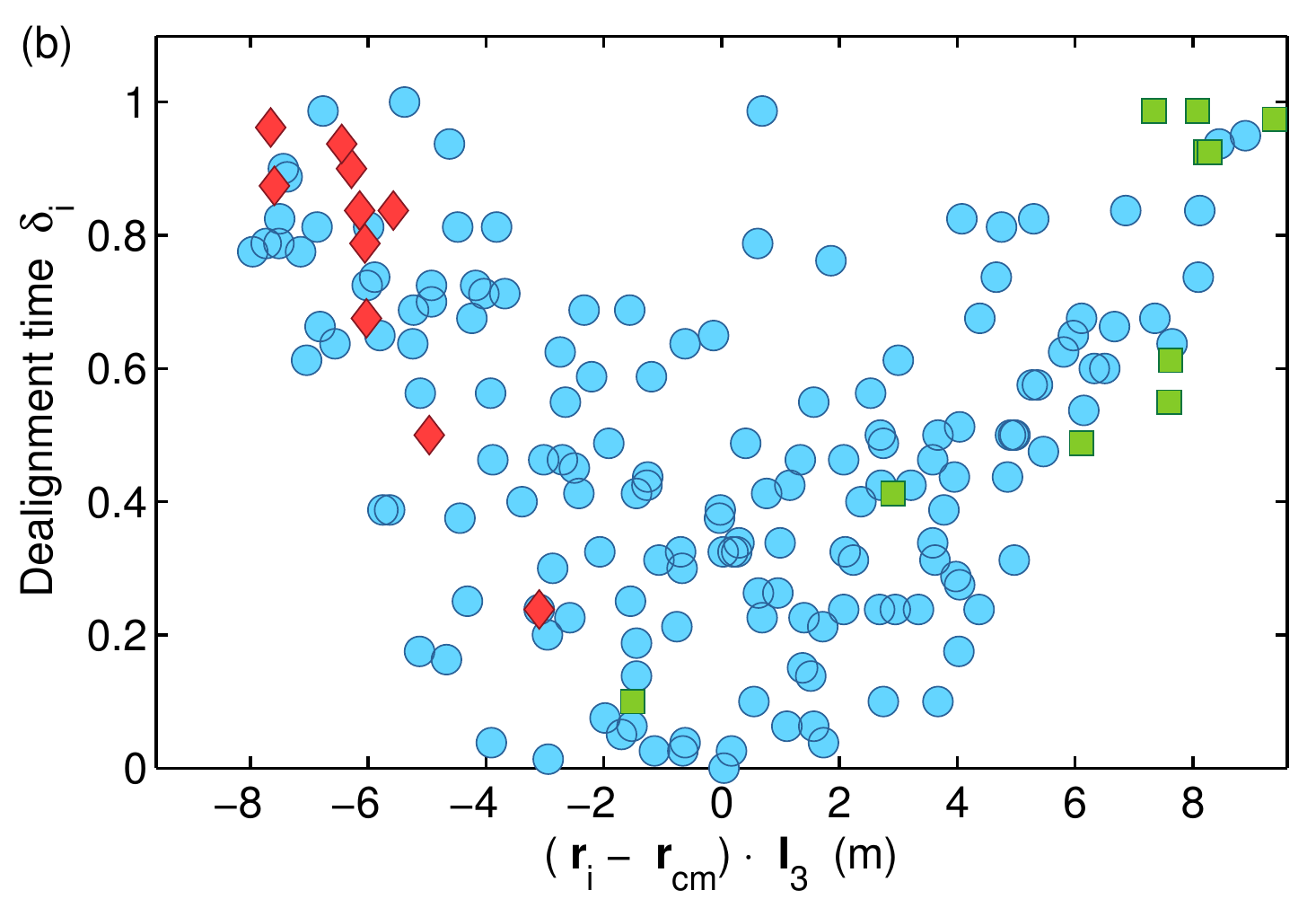}\\
\includegraphics[width=.95 \columnwidth]{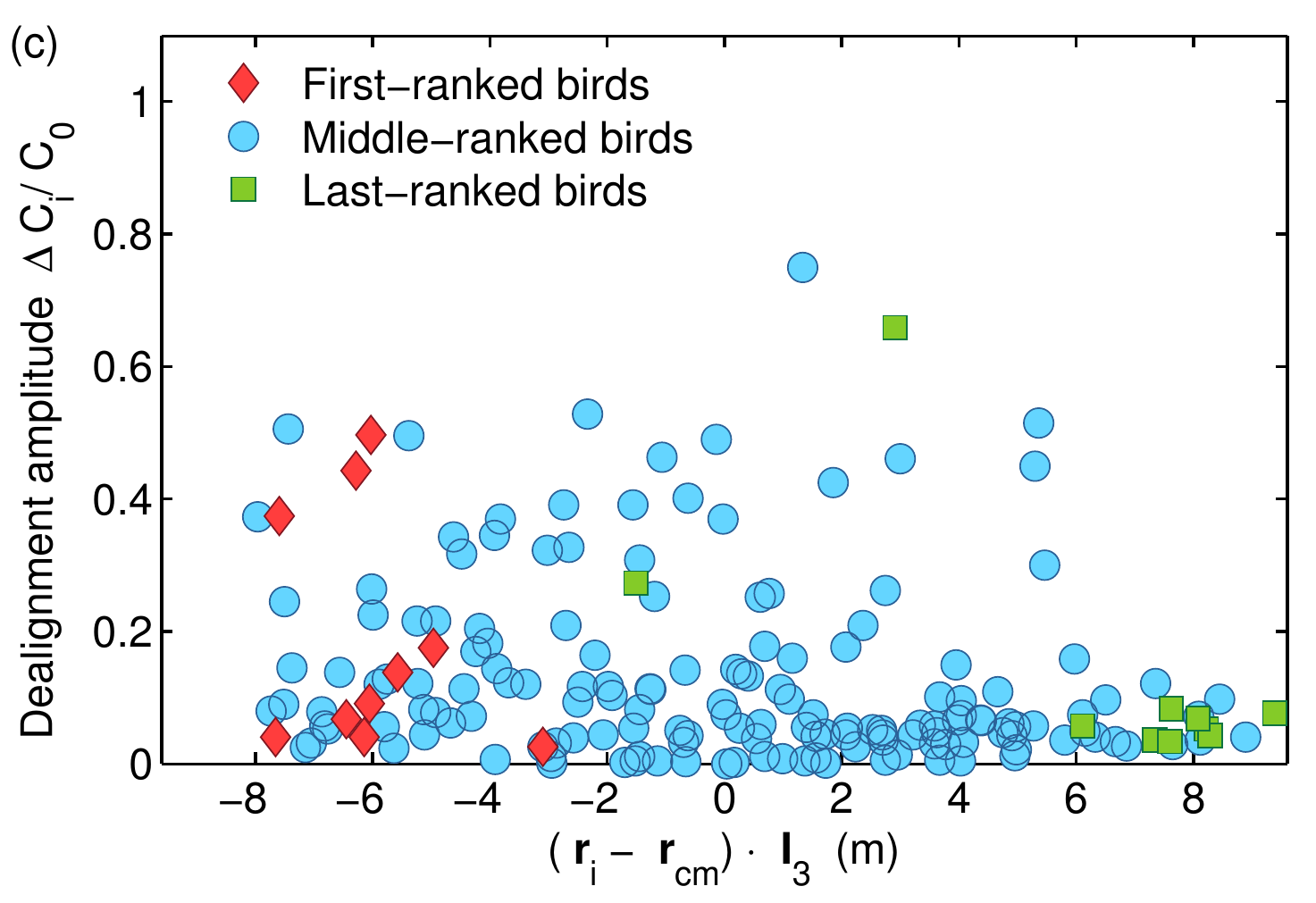}
 \caption{
{\bf Deviations from the global direction of motion.} 
(a) Directional correlation function $C_i(t)$ is plotted for three individual birds ranked $2$ (red), $60$ (blue), and $172$ (green) within a flock of $N=176$ birds performing a collective turn (event  E1 in Table.\ref{table:flocks}).
Only times prior and around the time of the start of the turn $t_1$ (vertical dashed line) are shown.
A median value of all $C_i(t)$ during time interval $\tau$ is used as a threshold value $C_0$ (horizontal dashed black line). 
(b) Dealignment time factor $\delta_i(\tau)$ is plotted as a function of position of bird $i$ along the longest elongation axis ${\bf I}_3$, with respect to the flock's center of mass. On average, the farther away from the center of the flock a bird is, the longer time its directional correlation is lower than $C_0$.
The top ten ranked birds (red diamonds) are situated close to one edge of the flock, while last ten birds to turn (green squares) are close to the other edge of the flock.
(c) Dealignment amplitude $\Delta C_i /C_0$ is shown as a function of position of bird $i$ along the longest elongation axis ${\bf I}_3$, with respect to the flock's center of mass. No obvious dependence on the position is observed.
 }
\label{fig:deviations}
\end{figure}

To this end, we define a directional correlation $C_i(t)$ between the direction of motion of an individual bird $i$ and the global direction of motion of the flock as
\begin{equation}
C_i(t) = \frac{{\bf v}_i(t)}{v_i(t)} \cdot \frac{{\bf V}(t)}{V(t)} \;.
\end{equation}
If at time $t$ bird $i$ is flying along the flock's global direction of motion, the value of the correlation is  $C_i(t)=1$, and we say that bird $i$ is completely aligned with the flock's direction.
However, birds are almost never completely aligned with the flock's direction of motion, at the very least due to wing-flapping (zig-zag of the trajectories can be seen in Video 1). Therefore, values of $C_i(t)$ fluctuate below, but close to $1$, while sometimes we observe strong decline of this quantity, when a bird deviates strongly from the global direction of motion (see Fig.\ref{fig:deviations}-a).

To understand which behaviour  might trigger the 
collective decision to make the turn, we analyze a time interval $\tau$ shortly before the start of the turn at $t=t_1$. During this interval we calculate the percentage of time during which bird $i$ deviates significantly from the global direction of motion. More specifically, we  check how frequently  its directional correlation $C_i(t)$ is lower than some threshold value $C_0$ (e.g. the median of all $C_i(t)$). We name this quantity a dealignment time factor $\delta_i(\tau)$ (see SI and Eq.\eqref{eq:dealign-time-factor} for precise definition).

In general, one might expect that all birds are deviating from a mean behaviour in a similar way, so that half of the time they fly more aligned to the global direction of motion (i.e.\ with $C_i(t)>C_0$), while other half of the time they deviate stronger from this direction (i.e.\ with $C_i(t)<C_0$). That would imply that the values for $\delta_i(\tau)$ are around $0.5$ for all the birds, independently of their position in the flock.
This is not however what happens. As can be seen in Fig.\ref{fig:deviations}-b and Fig.\ref{fig:deviations-SI-different-flocks} there is in fact a strong correlation between the location of an individual in the flock and the value of its dealignment factor: the farther away a bird is along the lateral elongation axis ${\bf I}_3$, the more frequently it exhibits  consistent fluctuations from the global direction of motion. On the contrary, no such correlation exists  when looking along the other elongation axes, or along the direction of motion (see Fig.\ref{fig:deviations-SI}).

The top ranked birds (red diamonds in Fig.\ref{fig:deviations}-b), which are situated close to one edge, are among the individuals with the  highest dealignment factor $\delta_i (\tau)$. We also computed the dealignment amplitude $\Delta C_i /C_0$, which gives a measure of how strongly bird $i$ deviates from the mean behavior during the time interval $\tau$ (see SI and Eq.\eqref{eq:dealign-amplitude}). While top ranked birds tend to have larger dealignment amplitude than average, they are not in general the individuals with strongest amplitude. There is not any clear correlation of the amplitude with the position along the elongation axis (see Fig.\ref{fig:deviations}-c). In fact, birds located  
centrally deviate as strongly or even stronger than the ones closer to the edges. However, their deviations are not followed by a global change of direction.  These results indicate that what really is distinctive in the behavior of initiators is how persistently they deviate from the flock's direction.  The turning event is triggered by the presence of such repeated deviations from average motion and is not due to a big, but momentary lapse of alignment.

Why do initiators, and edge individuals more generally, behave differently? 
Due to the asymmetric elongated shape of flocks (see Fig.\ref{fig:elongation} and Table \ref{table:flocks}), individuals in different locations can experience rather different boundary conditions. Birds in the bulk of the group are surrounded by many neighbors and well protected by external perturbations. They can momentarily fluctuate from the mean motion, but persistent large fluctuations are rare since volume confinement and social adaptation produce a strong feedback to the average. The same is not true for individuals at the border of the flock, and particularly for the ones on the extremal edges. These individuals are mostly surrounded by empty space. They can freely move towards the outside for a wide range of directions and experience an unbalanced social force by neighbors. They also are at large predatory risk, which might enhance their alertness and wish to relocate. All these factors contribute to produce persistent individual fluctuations, as signaled by the values of the dealignment time factor. Statistically this increases the probability of a coherent deviation of few individuals from the common flight direction, which might trigger spontaneous turns to occur.

\section{Discussion and Conclusions}
Spontaneous turns occur continuously during aerial display and flocks keep changing their direction of motion even in the absence of predators or obstacles. While responding to an external cue is a natural anti-predatory strategy, the occurrence of spontaneous collective maneuvers poses intriguing questions as to why and how this kind of behavior arises. In this work we quantitatively characterized collective turning and offered an explanation of their highly non-trivial dynamics. 

Our analysis, performed on  12 different turning events of various size and duration (see Table \ref{table:flocks}), outlines the following general scenario.

During straight flight flocks typically move level and acquire asymmetric shapes, with their longest elongation axis perpendicular to the flock's velocity.  This feature, that we clearly see in the data, has a well-known theoretical explanation: in self-propelled systems, orientational fluctuations perpendicular  to the global direction of motion (and to gravity) are larger than in other directions \cite{toner+tu_98,chate+al_08,cavagna+al_13a}, due to the breaking of the rotational symmetry in the polarized state. As a consequence,  a coherent group gradually elongates with its main axis  perpendicular to its velocity and to gravity (see Fig.\ref{fig:elongation}-c and Table \ref{table:flocks}). One of the main outcomes of this process is that birds situated in different regions of the flock experience rather different boundary conditions. In particular, the individuals in the elongated edges of the group have much less spatial confinement, and a strongly anisotropic distribution of neighbors. These conditions cause a larger directional mobility, and in fact  our data show that individuals at the edges are the ones that more persistently deviate  from the global direction of motion.  For some of these birds this effect can become so strong as to compensate the feedback to average motion, and initiate a turn. Once the turn starts - locally at the lateral edges - it propagates through the group thanks to the social interactions between individuals. Birds turn following equal radius paths and rearrange their positions with respect to the global flock's direction, so that at the end of the turn the very individuals that occupied lateral positions are in the front/back of the group, and the whole flock has its longest elongation axis oriented along the direction of motion.

There are a few interesting aspects of this collective phenomenon, which we would like to comment further. 

One of our main findings is that turns systematically start on the outer tips of the flock. The fact that individuals on the border of a group behave differently has been already discussed in the literature on collective animal behavior. Staying on the border is not usually a preferential location during collective motion, since these positions suffer higher risk under a predator attack or any other external perturbation. Therefore one might expect border individuals to be particularly  risk alert and exhibit a strong pressure to exchange position for a more favorable one.  
Several studies have been performed on feeding groups. While feeding on the ground, birds usually stay in flocks. Observations on starlings and other birds species show that individuals on the edges of the flock display stronger vigilance, that is, they scan the surrounding more frequently for predators and they feed more rapidly, etc. (see e.g.\cite{lazarus_78,jennings_80,inglis_81,black_92,fernandez_08} ). 
In the case of flying turning flocks, we find that border individuals deviate from the mean motion more frequently than others. It might be that due to their location these birds increase their alertness so that each of them becomes more sensitive and prone to changes, much as border individuals in feeding groups have a larger scanning and feeding rate. However this is not the only possible cause for their anomalous behavior. We know that when moving collectively birds coordinate with one another using local adaptation rules (i.e. `align with your neighbors') \cite{aoki_82,vicsek+al_95,couzin+al_02,hemelrijk+al_11,bialek+al_12}. In this respect, being close to the edges implies having an atypical neighborhood, thereby experiencing a social force significantly different than inside the flock. This very fact can produce more persistent fluctuations at the border and trigger turns. 
 While it is certainly very difficult to disentangle these effects in real data, the idea that social positional heterogeneities are important to explain the role of edge individuals is in our opinion worth investigating in future studies.

The fact that top ranked birds display unusual directional fluctuations is also interesting within another respect. We find that what really matters to initiate a turn is that these birds deviate from average motion much more frequently than others. On the other hand, we find no correlation between the strength of such behavioral deviations and turning (Figs.~\ref{fig:deviations}-c and \ref{fig:deviations-SI-different-flocks}).
These results indicate that the response of the neighbors to the initiator of the turn is related to a repeated alarm signal--perceived as a deviation from the average motion--exhibited by the initiator, and not to a sudden and strong fluctuation. Similar results have been found in the works concerning vigilance behavior in feeding birds, where it has been shown that response to alarm cues occurs with repeated detection of an alarm signal \cite{cresswell_00, lima_94}.  It has been argued that this strategy is useful for avoiding false alarms which are common, since taking flight in the course of feeding is both costly in energy and time 
(see e.g. \cite{proctor_01}).


 As we discussed so far, turns can and in fact do  occur spontaneously. One can wonder  why this is the case, and to what extent such a collective behavior can be advantageous to the individuals. On the one hand, spontaneous turns are a social response of the flock to the persistent deviations of some edge individuals. We can expect a similar mechanism to occur when an external repeated signal, like an approaching predator or an obstacle, is encountered. In this sense, spontaneous turns are a byproduct of the very ability of the group to respond collectively to local perturbations. On the other hand, our results show that during a turn individuals change their positions with respect to motion, with the remarkable consequence that birds who suffered extremal conditions before the turn acquire better locations after it. Risk is redistributed between individuals at no expense of global order. In this respect, 
the whole process of turning is a remarkable example of how a self-organized system can sustain collective changes and reorganize, while retaining coherence.

%
%

%

\section{Acknowledgments}

This work was supported by grants IIT--Seed Artswarm, ERC--StG n.257126 and US-AFOSR - FA95501010250 (through the University of Maryland).




\newpage

\onecolumngrid

\vfill\eject

\centerline{ \bf  SUPPLEMENTARY INFORMATION}


\setcounter{section}{13}  

 \renewcommand{\theequation}{S\arabic{equation}}
 \setcounter{equation}{0}  
 
\renewcommand{\thefigure}{S\arabic{figure}}
 \setcounter{figure}{0}  

\renewcommand{\thetable}{S\arabic{table}}
 \setcounter{table}{0}  

\section*{Methods}

\noindent
{\bf Experiments}.
During winter, flocks of European starlings ({\it Sturnus vulgaris}) are a common sight in Rome, where they populate several roosting sites. We collected data shortly before dusk at the site of Piaz\-za dei Cinquecento, between November 2010 and December 2012. The video sequences are acquired using the trifocal stereometric setup described in ref.~\cite{cavagna+al_08a}. Two cameras separated by a baseline distance $D_{12}=25$m are the stereometric pair. A third camera, placed at a shorter distance $D_{23}=2.5$m from the first one, allows us to exploit the trifocal constraint for solving the stereo correspondence (matching) problem  \cite{cavagna+al_08a}. We employ three high--speed cameras IDT-M5 with monochromatic CMOS sensor with resolution $2288\times1728$ pixels, shooting at $170$ fps. Cameras store images on off-board memory using the Camera Link protocol. Lenses used are Schneider Xenoplan $28$mm $f/2.0$. Typical exposure parameters are: aperture between $f/2.8$ and $f/8$; exposure time between $700$ and $3500$ms. Intrinsic camera parameters are calibrated every two weeks in the laboratory using a set of $50$ images of a planar target. The accuracy of the 3D system is regularly tested using laser-metered artificial targets. Flocks perform turns typically at a distance of $80$-$130$m from the cameras. The error on the relative distance between two neighbouring birds is $\sim0.1$m. Time duration of the recorded events is between $1.8$ and $12.9$s. 
We reconstructed the 3D positions and velocities of individual birds using the techniques developed in ref.~\cite{attanasi+al_13b}. 
The data set consists of 12 distinct flocking events, each one including one collective turn, as reported in Table  \ref{table:flocks}. If on the recorded sequence the flock performs more than one turn, the time lag is chopped and different turns are studied as independent events (e.g.\ events E11 and E12 in Table \ref{table:flocks}). 

\begin{table*}[h!]
\vskip 0.1 in
\footnotesize
\begin{tabular}{clcccccccccc}
\hline 
\hline 
 {\sc Event }  &
 \hspace{0.5cm}{\sc Event }  & 
 \hspace{0.2cm}$N$  \hspace{0.2cm}& 
 \hspace{0.2cm} $\Phi$      \hspace{0.1cm}   & 
  \hspace{0.1cm}    $ {\bf I}_1 \cdot {\bf G} $    \hspace{0.1cm} &
  \hspace{0.1cm}   $ {\bf I}_3 \cdot {\bf G} $      \hspace{0.1cm} &
  \hspace{0.1cm}   $ {\bf V}_1 \cdot {\bf G} $    \hspace{0.1cm} &
    \hspace{0.1cm} $ {\bf I}_3 \cdot {\bf V}_1$  \hspace{0.1cm}   &
   \hspace{0.1cm}  $ {\bf I}_3 \cdot {\bf V}_2$    \hspace{0.1cm} &
  \hspace{0.1cm}   $ {\bf I}_3 \cdot \frac{{\bf d}_0}{\|{\bf d}_0\|}$    \hspace{0.1cm}  &  
  \hspace{0.cm}   $ {\bf V}_1 \cdot \frac{{\bf d}_0}{\|{\bf d}_0\|}$  \hspace{0.cm}   &
    \hspace{0.1cm}   $ {\bf V}_2 \cdot \frac{{\bf d}_0}{\|{\bf d}_0\|}$  \hspace{0.cm}     
 \\
   {\sc number}  & 
  \hspace{0.6cm}{\sc label}  & 
  &
  &
    at $t_1$   &
   at $t_1$    &
   at $t_1$    &   
   at $t_1$   & 
   at $t_2$   &
   at $t_1$    &   
   at $t_1$     & 
\\
 \hline
  \hline
E1& 20110208\_ACQ3 & 176 & 0.806 &  0.95 & 0.15 & 0.47 & 0.14 & 0.93 & 0.97 & 0.08 & -0.91 \\
E2& 20111124\_ACQ1 & 125 & 0.959 &    0.97 & 0.15 & 0.01 & 0.65 & 0.99 & 0.97 & 0.81 & -0.78 \\
E3& 20111125\_ACQ1 & 50 & 0.866 &   0.86 & 0.42 & 0.26 & 0.41 & 0.88 & 0.91 & 0.05 & -0.82 \\
E4& 20111214\_ACQ4\_F1 & 154 & 0.940 &  0.69 & 0.72 & 0.03 & 0.12 & 0.63 & 0.97 & 0.10 & -0.83 \\
E5& 20111215\_ACQ1 &  384 & 0.801 &   0.97 & 0.16 & 0.13 & 0.24 & 0.92 & 0.98 & 0.05 & -0.94 \\
E6& 20111125\_ACQ2 &  502 & 0.841 &  0.98 & 0.16 & 0.35 & 0.38 & 0.89 & 1.00 & 0.43 & 0.74 \\
E7& 20110217\_ACQ2 &  404 & 0.854  &  0.91 & 0.32 & 0.39 & 0.34 & 0.96 & 0.78 & 0.83 & -0.89 \\
E8& 20111220\_ACQ2 &  197 & 0.907   & 0.98 & 0.03 & 0.94 & 0.30 & 0.90 & 0.56 & 0.20 & -0.53 \\
E9& 20111201\_ACQ3\_F1 &  133 & 0.793   &  0.76 & 0.46 & 0.04 & 0.21 & 0.99 & 0.89 & 0.21 & 0.78 \\
E10 & 20110211\_ACQ1 &  595 & 0.757  &  0.94 & 0.10 & 0.34 & 0.03 & 0.36 & 0.91 & 0.37 & -0.96 \\
E11 & 20111214\_ACQ4\_F2\_T1&  139 & 0.890  & 0.35 & 0.87 & 0.67  & 0.92 & 0.56 & 0.84 & 0.74 & 0.08 \\
E12 & 20111214\_ACQ4\_F2\_T2 &  139 & 0.808 & 0.66 & 0.74 & 0.11 & 0.58 & 0.79 & 0.89 & 0.15 & 0.39 \\ 
\hline
\hline 
\end{tabular}
\caption{{\bf Global quantitative properties of the turning events.}
We analyzed twelve turning events, of which two (E11 and E12) are performed by the same flock one after the other (therefore marked T1 and T2 in the event label).
$N$ is the number of birds in the flock. 
The polarization is defined as $\Phi = \|(1/N) \sum_i {\bf v}_i/\|{\bf v}_i \| \|$.
In the remaining columns we report absolute values of the scalar products between 
yaw ${\bf I}_1$ (the axis relative to the shortest dimension of the flock), 
the longest elongation axis ${\bf I}_3$, 
and gravity ${\bf G}$, 
with the direction of motion before and after the turn 
given by velocity vectors ${\bf V}_1$ and ${\bf V}_2$ at times $t_1$ and $t_2$, respectively.
The scalar products are calculated using the values of appropriate quantities at times $t_1$ of the 
start of the turn, or $t_2$ at the end of the turn, as indicated.
Note that the vectors ${\bf I}_1$, ${\bf I}_3$, and ${\bf G}$ are unitary by definition, while for the
direction of motion we used normalized velocity vectors 
${\bf n}_1 \equiv  {\bf n}(t_1)= {\bf V}(t_1)/\|{\bf V}(t_1)\|$ and
${\bf n}(t_2) = {\bf V}(t_2)/\|{\bf V}(t_2)\|$,
which are for clarity called ${\bf V}_1$ and ${\bf V}_2$ in the column titles.
Finally, in order to quantify the location of the origin of the turn, we use a mean relative position of the $10$ top-ranked birds with respect to the barycenter of the flock, ${\bf d}_0$, as defined in the main text. 
We calculate absolute scalar products of the normalized vector ${\bf d}_0/\|{\bf d}_0\|$ with the unitary vector ${\bf I}_3$ of the longest elongation axes at $t_1$, as well as with the direction of motion at the start of the turn ${\bf n}_1$.
In the last column, we report the scalar product between ${\bf d}_0/\|{\bf d}_0\|$ and the new direction of motion after the turn, given by the unitary velocity vector ${\bf n}(t_2)={\bf V}(t_2)/\|{\bf V}(t_2)\|$ (called ${\bf V}_2$ for simplicity).
Note that absolute values of all scalar products are reported, except for the last one whose negative values signify that the top-ranked birds initiated the turn in the direction towards the flock's barycenter (towards inside of the flock and not outside).
\label{table:flocks}}
\end{table*}

\newpage

\section*{Reorientation during turns}

\subsection*{Global orientation of the flock}

\noindent
{\bf Turning plane.}
It has previously been observed that the trajectories of birds during a turn lie approximately on a plane (see ref.~\cite{attanasi+al_14} and Fig.\ref{fig:turning-plane}-a). We exploit this fact in order to simplify the representation of a turn propagation through the flock and subsequent analysis. 
We determine the `turning plane'  by using two average flock velocity vectors: ${\bf V}_1 \equiv {\bf V}(t_1)$ at time $t=t_1$ of the start of the turn, and ${\bf V}_2 \equiv {\bf V}(t_2)$ at time $t=t_2$ shortly after the turn is finished.
The average flock velocity at time $t$ is calculated as ${\bf V}(t) =  \frac{1}{N} \sum_{i=1}^{N}{\bf v}_i(t)$, where ${\bf v}_i(t)$ is the velocity of bird $i$ at time $t$. The `turning plane' can then be defined by its normal unit vector ${\bf n}_3$ which is perpendicular to the plane, and the flock's barycenter position at time $t_1$. Vector ${\bf n}_3$ is obtained as a unit vector orthogonal to ${\bf V}_1$, as well as to ${\bf V}_2$, that both lie within the `turning plane', that is 
${\bf n}_3 = ({\bf V}_1\times {\bf V}_2 )/ \|{\bf V}_1\times {\bf V}_2 \| $. Finally, the normal to the plane ${\bf n}_3$, together with a unit vector ${\bf n}_1= {\bf V}_1/\|{\bf V}_1\|$ along the direction of motion at the start of the turn $t_1$, and a unit vector ${\bf n}_2 = {\bf n}_3 \times {\bf n}_1$, form an orthogonal coordinate system  $({\bf n}_1,{\bf n}_2,{\bf n}_3)$ (see Fig.1-a). 
We typically use this coordinate system $({\bf n}_1,{\bf n}_2,{\bf n}_3)$ for data visualization in the paper.
Note that times $t_1$ and $t_2$ are not necessarily initial and final times of the acquisition, but the times before and after the flock turns. They are estimated either from the trajectories themselves, or from the calculated velocities and radial accelerations. 

\begin{figure}[b] 
  \centering
\includegraphics[width=0.32\columnwidth]{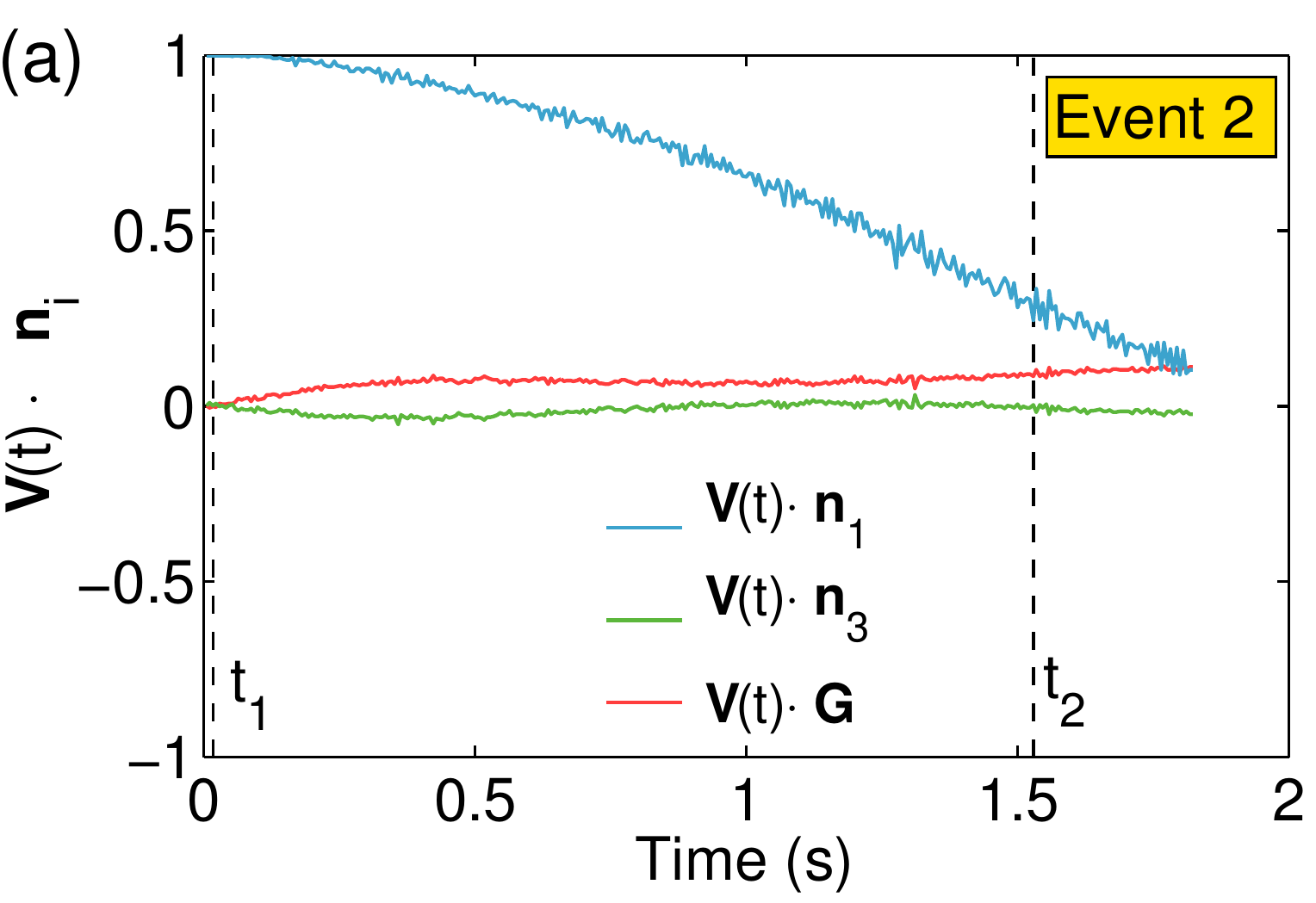}
\includegraphics[width=0.32\columnwidth]{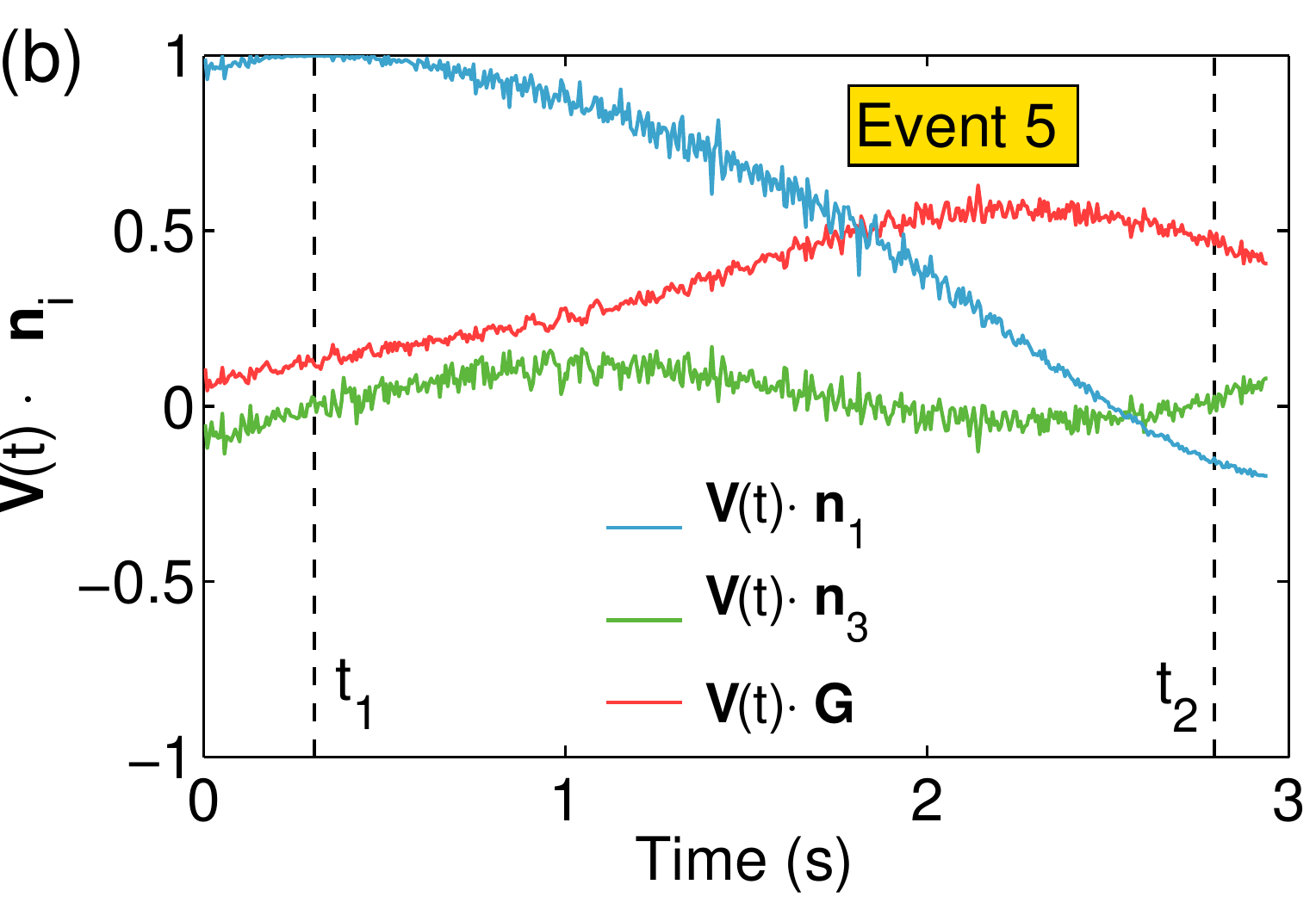}
\includegraphics[width=0.32\columnwidth]{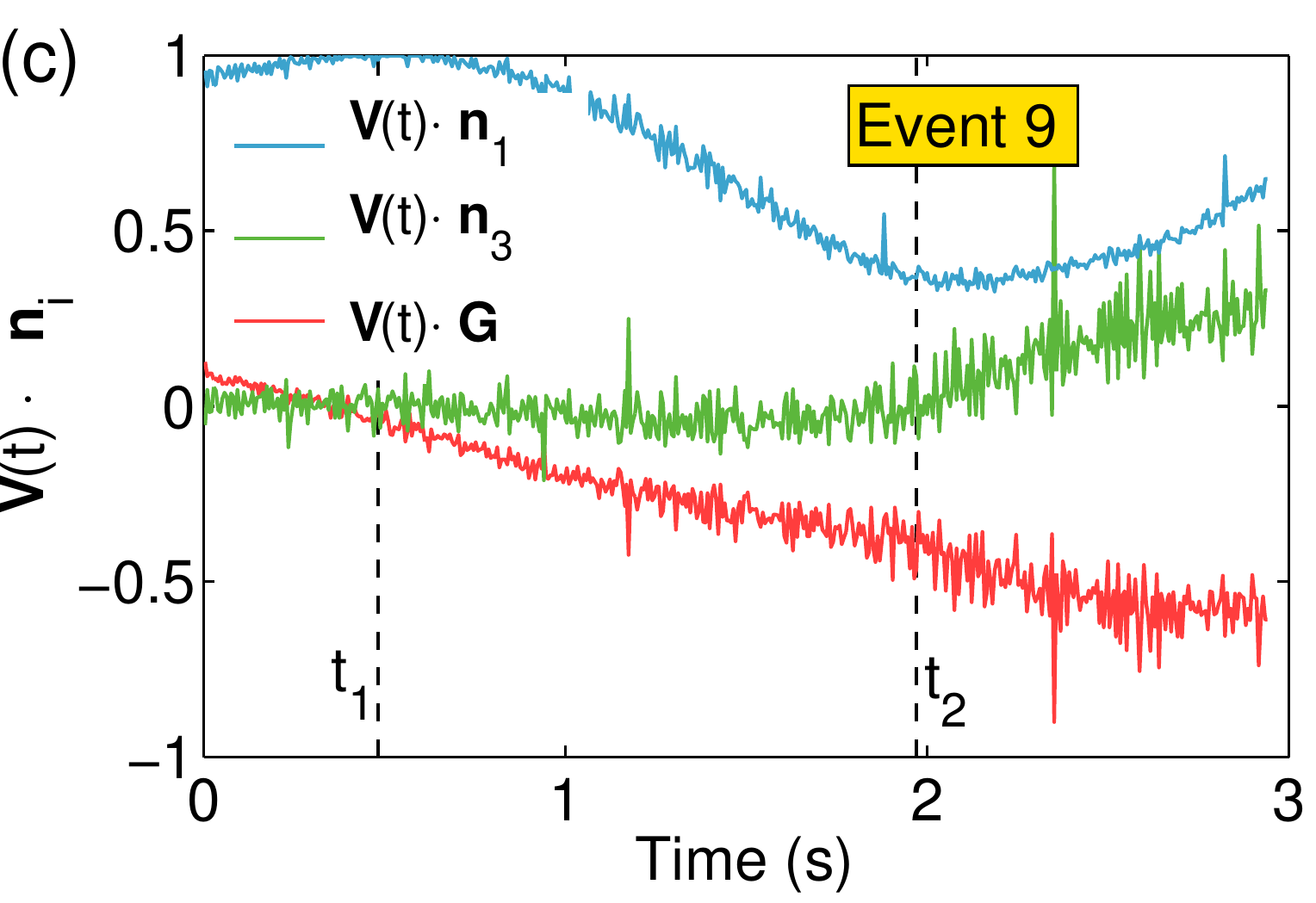}\\
\includegraphics[width=0.32 \columnwidth]{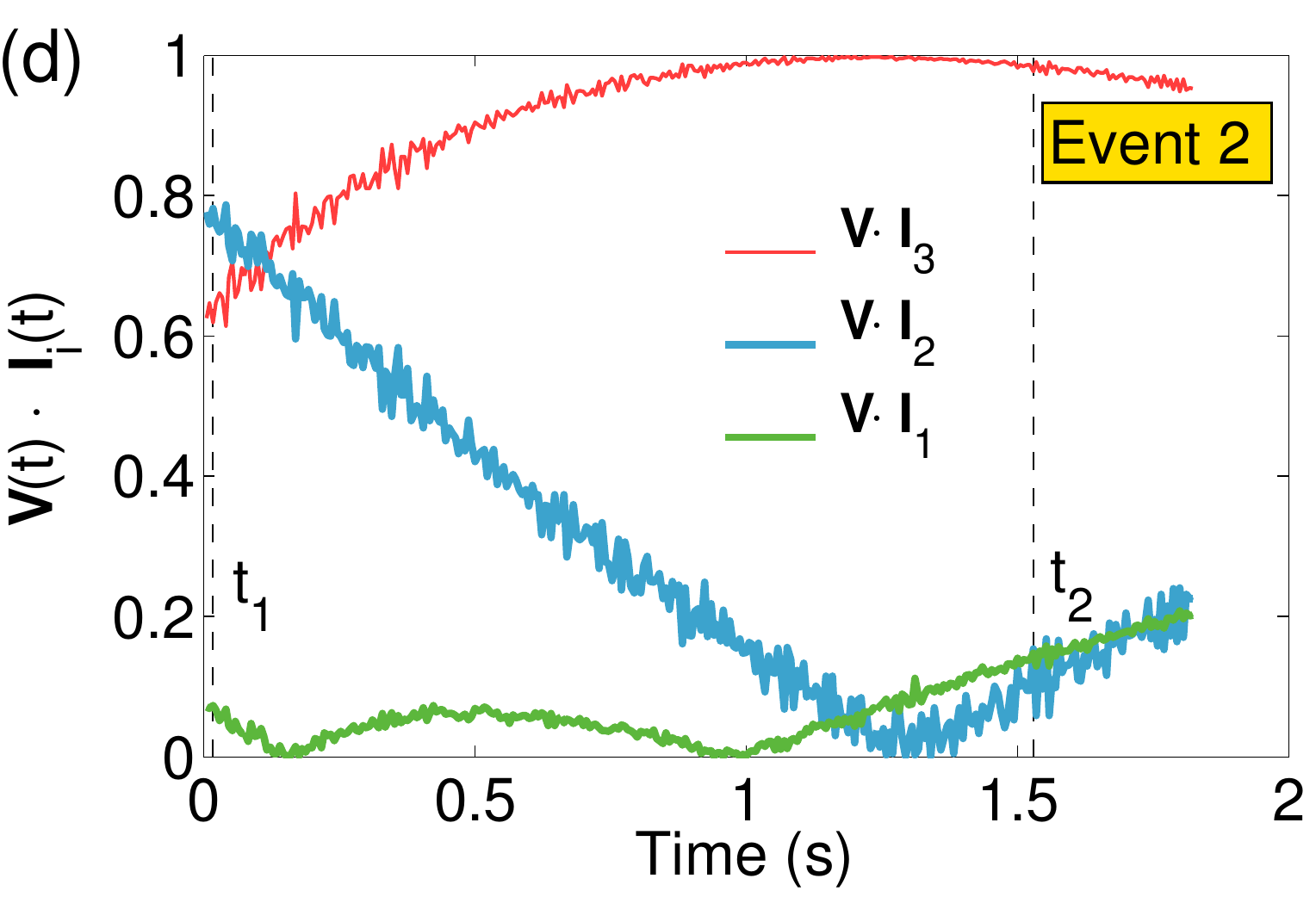}
\includegraphics[width=0.32 \columnwidth]{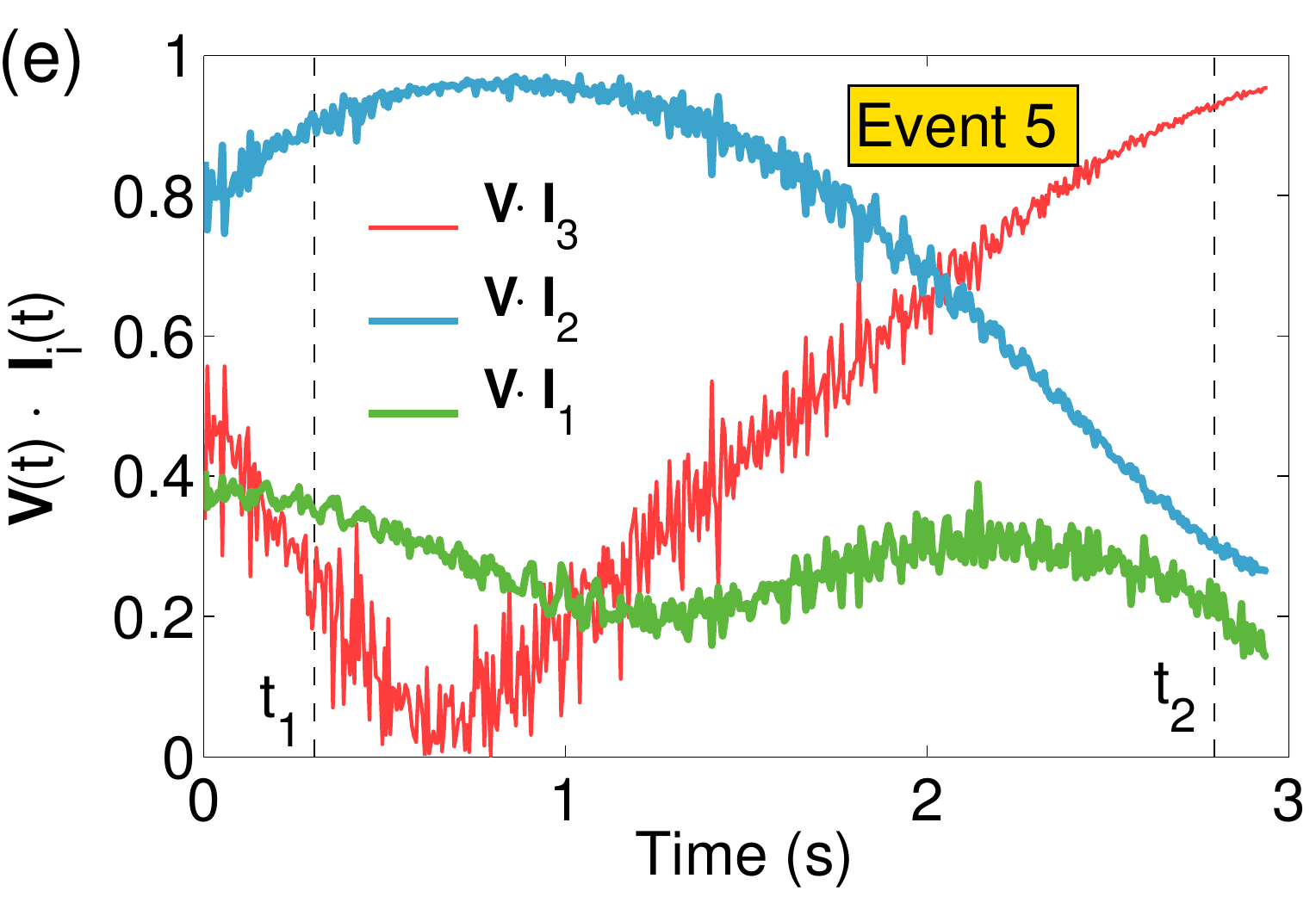}
\includegraphics[width=0.32 \columnwidth]{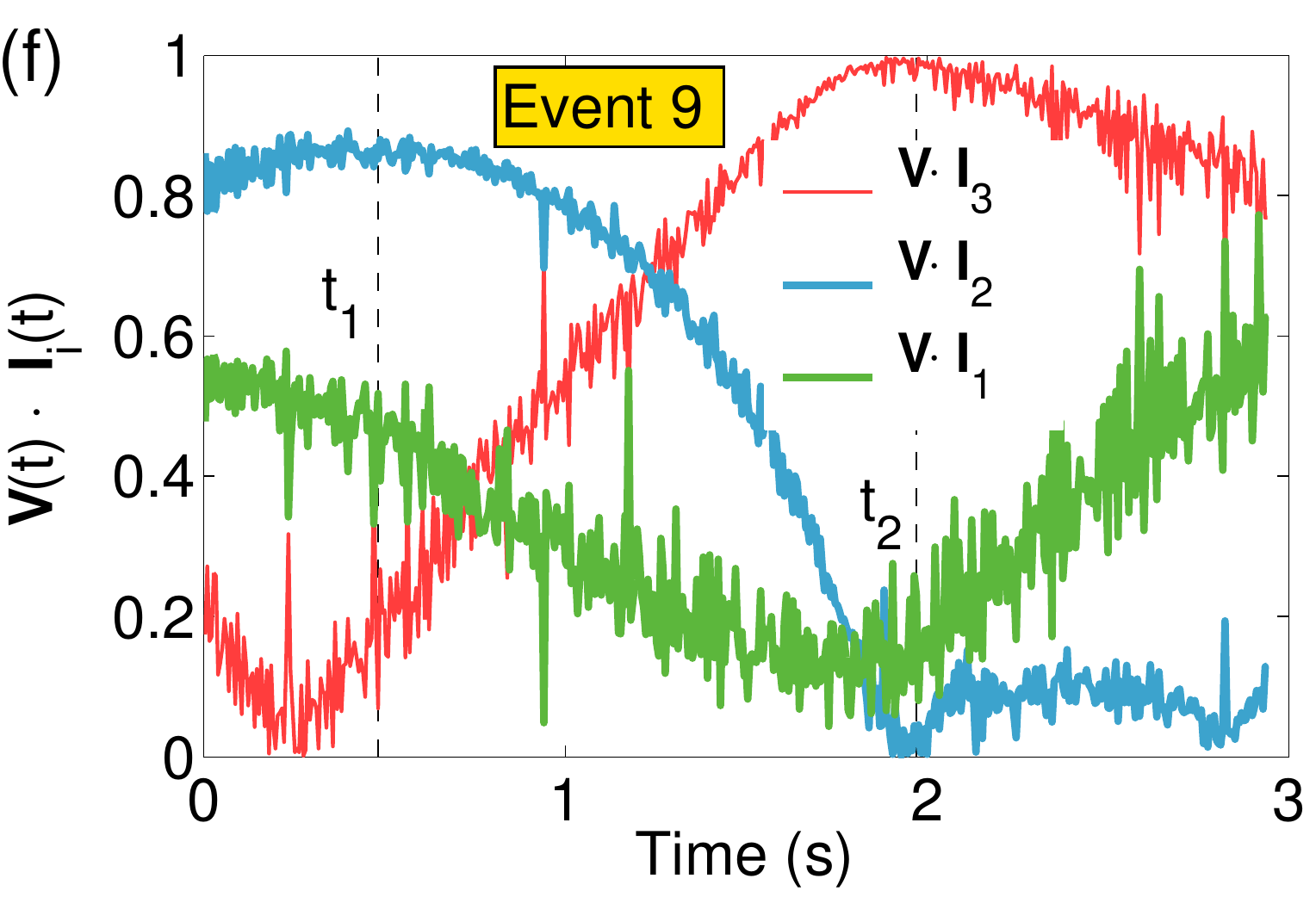}
 \caption{
{\bf Flock's direction of motion and orientation during turning events.} 
For three turning events labelled E2, E5, and E9 (for details see Table \ref{table:flocks}), we study the change in the flock's direction of motion and its orientation in space during a turn. Times: $t=t_1$ of the start of the turn, and $t=t_2$ after the turn are marked by vertical dashed lines. 
Panels (a), (b), (c) on the top show how the flock's global direction of motion, given by a unit vector of its velocity ${\bf V}(t)$, changes over time with respect to three characteristic directions: 
i) direction of motion at the start of the turn $t_1$ given by a unit vector ${\bf n}_1$, 
ii) normal to the `turning plane'  ${\bf n}_3$, and 
iii) gravity given by a unit vector ${\bf G}$.
Scalar product ${\bf V}(t)\cdot{\bf n}_1$ (blue curve) quantifies how much the direction of motion changes during the turn: it is $1$ at time $t_1$, while its final value corresponds to the angle between the initial and final direction of motion. The value of the scalar product ${\bf V}(t)\cdot{\bf n}_3$ (green curve) fluctuates around $0$ during the turn between times $t_1$ and $t_2$, confirming that the turn lies on a `turning plane' given by its normal vector ${\bf n}_3$. Finally, scalar product ${\bf V}(t)\cdot{\bf G}$ (red curve) shows whether the flock is turning parallel to the ground (values around $0$) or not.
Panels (d), (e), (f) on the bottom show the evolution of the relation between the flock's global direction of motion (given by its velocity unit vector ${\bf V}(t)$) and its elongation axes ${\bf I}_{1}$, ${\bf I}_{2}$, and ${\bf I}_{3}$ over the time of the turn. 
}
\label{fig:velocity-scalar-prod}
\end{figure}

During the turn, between $t_1$ and $t_2$, the flock turns on the above defined turning plane. This can be verified by looking at the scalar product between the unit vector of velocity (representing direction of motion) and the normal to the plane ${\bf n}_3$. Indeed, in Fig.~\ref{fig:velocity-scalar-prod} one can see that this scalar product, ${\bf n}_3\cdot{\bf V}(t)/V(t)$, fluctuates around zero, thus confirming the 'turning plane' observation. Other scalar products of interest are also given, such as: ${\bf V}(t)\cdot{\bf n}_1$ that quantifies the change of the flock's direction of motion over time with respect to the start of the turn at $t=t_1$, and ${\bf V}(t)\cdot{\bf G}$ (with ${\bf G}$ being gravity unit vector) that shows whether a flock is flying parallel to the ground or not.

\vspace{0.4 cm}
\noindent
{\bf Elongation axes.}
It is convenient to consider the three principal elongation axes of the flock ${\bf I}_1$, ${\bf I}_2$, and ${\bf I}_3$. Since flocks typically have asymmetric flat shapes the dimensions of the group along these axes are $I_1< I_2<I_3$. The smallest dimension $I_1$ characterizes the thinness of the flock and its relative axis ${\bf I}_1$ defines the yaw axis of the flock. Flocks typically fly parallel to the ground with their yaw axis parallel to gravity (see \cite{ballerini+al_08b} and Table \ref{table:flocks}). To study the re-orientation of the flock during the turn, we computed the angles between the three axes and the flock's velocity and checked their evolution in time from the start to the end of the turn (see Fig.~\ref{fig:velocity-scalar-prod}). Since turns are almost planar, a synthetic description of this dynamics is given by  the normalized scalar product of the longest elongation axis ${\bf I}_3$ and the flock's velocity before and after the turn, i.e.\ at times $t_1$ and $t_2$ for all our turning events. 
The result is displayed in Fig.\ref{fig:elongation}, and confirms what is qualitatively observed in  Fig.\ref{fig:turn-propagation-rainbow}. 
The values of these scalar products reported in Table \ref{table:flocks} confirm that prior to the turn the flock's longest axis tends to be perpendicular to its direction of motion (small values ${\bf I}_3 \cdot {\bf V}$ at $t_1$), while during the turn it aligns more with the flock's velocity (high values ${\bf I}_3 \cdot {\bf V}$ at $t_2$).  

Typically, the values of the scalar product ${\bf I}_3 \cdot {\bf V}$ at the start of the turn $t_1$ are below $0.5$, corresponding to an angle between $60^{\rm o}$ and  $90^{\rm o}$ between the longest axis and velocity. There are a few exceptions to this general conclusion, which can be easily explained. 
On the one hand, the exceptions are events E11 and E12 which correspond to two consecutive turns of the same flock,
and occur after a merging of two separate flocks into one (as revealed by the original videos of the events), which might be the reason  why the reorientation is different. 
In addition, for these two events we find that the longest elongation axis ${\bf I}_3$ is more aligned to gravity ${\bf G}$ than the yaw axis ${\bf I}_1$, contrary to all other flocks we analyzed (see Table \ref{table:flocks}) and results in \cite{ballerini+al_08b}. 
On the other hand, for event E2, the data acquisition initiated too late to capture the very start of the turn. Although this did not hinder the determination of the ranking curve,  the orientation of the flock already started to change prior to the start of the acquisition, preventing us from obtaining the initial orientation angle.

%
%

\newpage
\subsection*{Propagation during turns}

\begin{figure}[h]
  \centering
\includegraphics[width=0.9\columnwidth,]{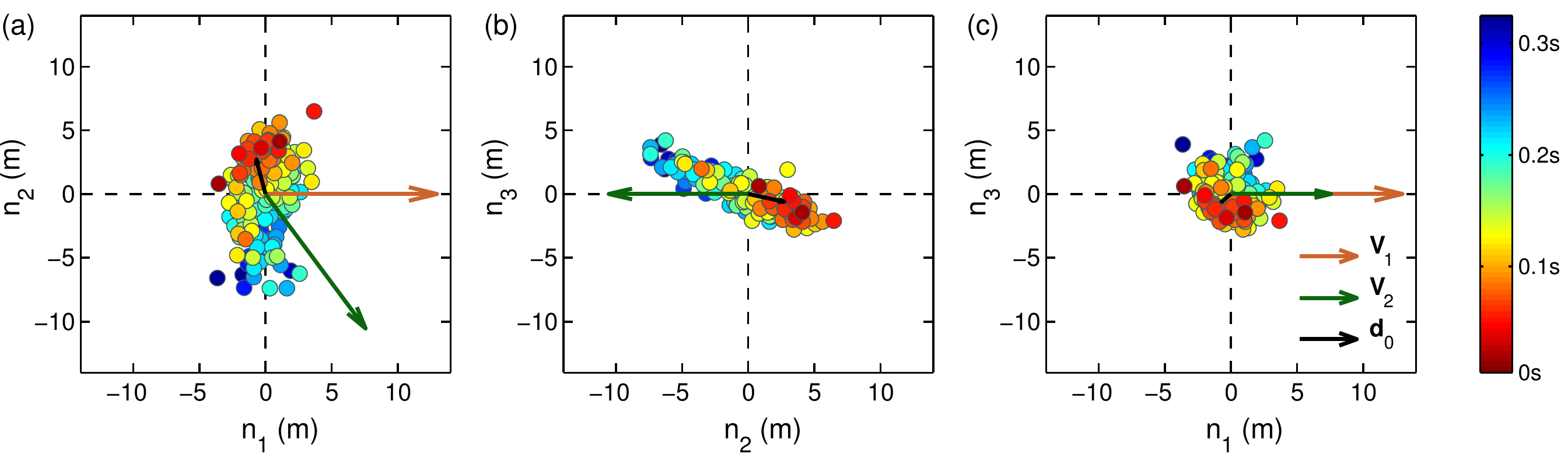}\\[0.5cm]
 \includegraphics[width=0.9\columnwidth,]{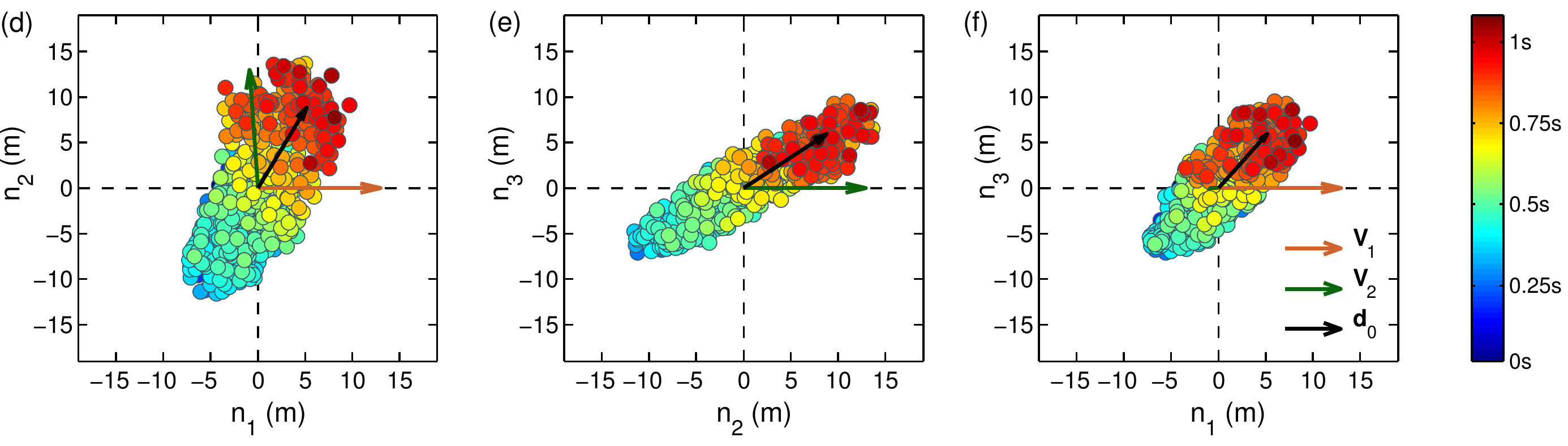}\\[0.5cm] \includegraphics[width=0.9\columnwidth,]{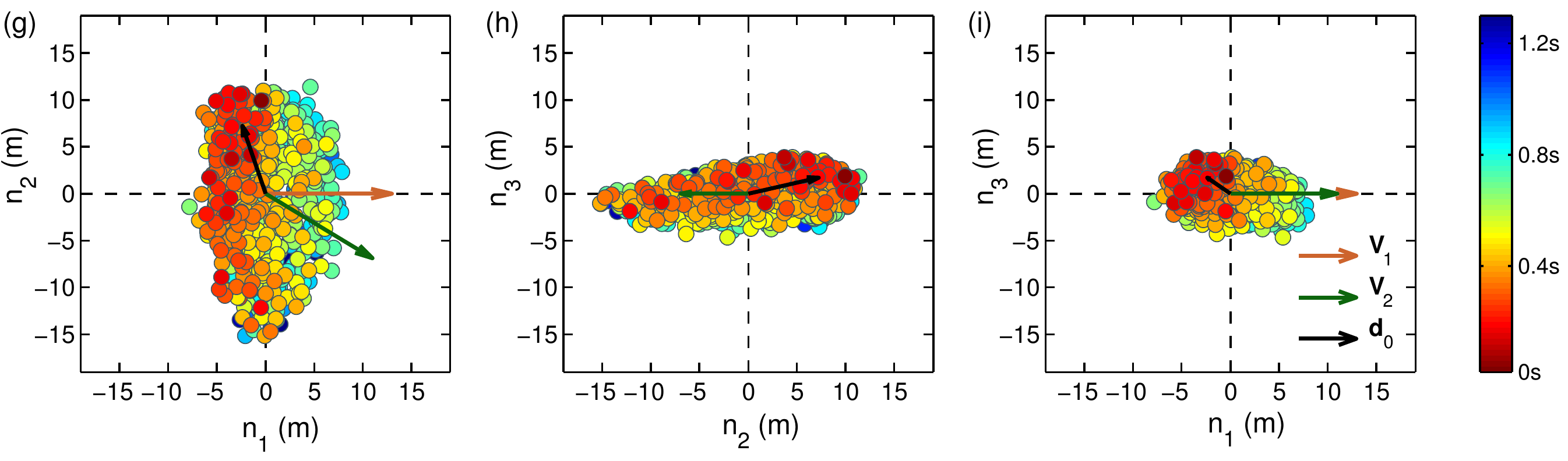}
 \caption{{\bf Propagation of the turn through a flock.} 
We show propagation of the turn through a flock for several different turning events: event number E4 in top panels; event number E6 in middle panels; event number E10 in bottom panels.
The flocks are shown at moment  $t_1$ of the start of their turn 
by using the projections on the planes of the `turning plane' orthogonal system $({\bf n}_1,{\bf n}_2,{\bf n}_3)$ centered at the flock's center of mass. 
Different panels show a view on the flocks (a), (d), (g) from the top, (b), (e), (h) from the front, and (c), (f), (i) from a side.
At time $t_1$ each flock is moving along the direction $n_1$ with velocity ${\bf V}_1$, which is to the right in panels (a), (d), (g) and (c), (f), (i), while in panels (b), (e), (h) they move towards the viewer. At moment $t_2$ after the turn, each flock is moving with the velocity ${\bf V}_2$. The velocity vectors are rescaled (increased), in order to emphasize the old and new direction of global flock's movement. 
A wave of directional change information spreads through each of the flock during a turn. The birds are colored according to their turning time delays $t_i$, as indicated by the colorbar on the right. The mean position of the first $5$ birds to turn at time $t_1$, with respect to the barycenter, is indicated by vector ${\bf d}_0$. 
 }
\label{fig:turn-propagation-rainbow-3flocks}
\end{figure}

In Fig.~\ref{fig:turn-propagation-rainbow-3flocks} we show how the information to change the direction of motion  propagates from the first bird through an entire flock. In the same way as in Fig.~\ref{fig:turn-propagation-rainbow}, we  project the birds' positions at the start of the turn to the `turning plane' coordinate system $({\bf n}_1,{\bf n}_2,{\bf n}_3)$ centered at the flock's barycenter at time $t_1$. The color used for each bird reveals its turning time delay $t_i$ with respect to the first bird to turn to which we assign the delay $t_1=0$~s. The values of the time delays are used from the established ranking in ref.~\cite{attanasi+al_14}.

\newpage
\subsection*{Equal radius paths and network rearrangement}

\begin{figure}[h] 
\centering
 \includegraphics[width=0.42 \columnwidth]{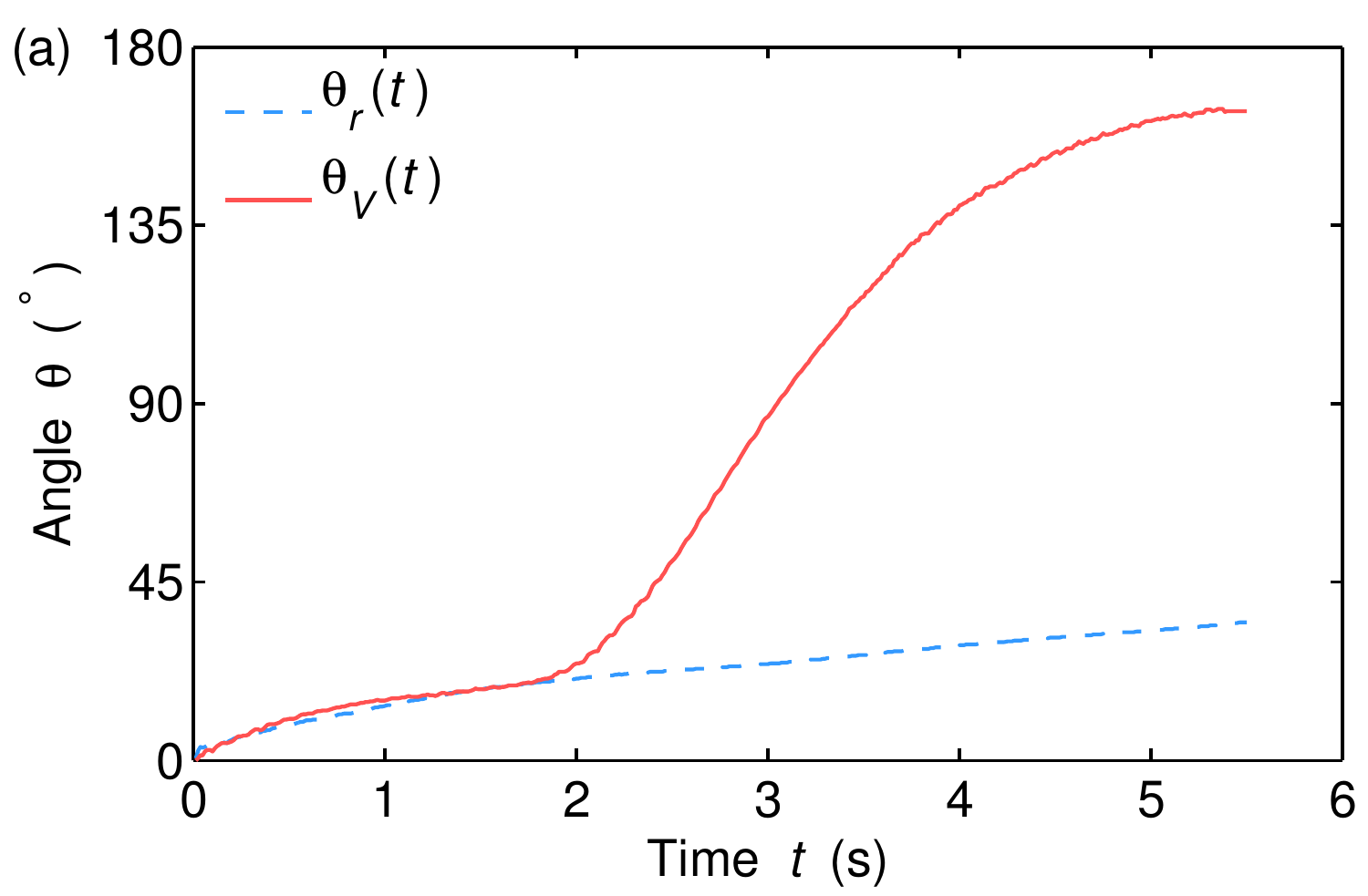} 
 \includegraphics[width=0.42 \columnwidth]{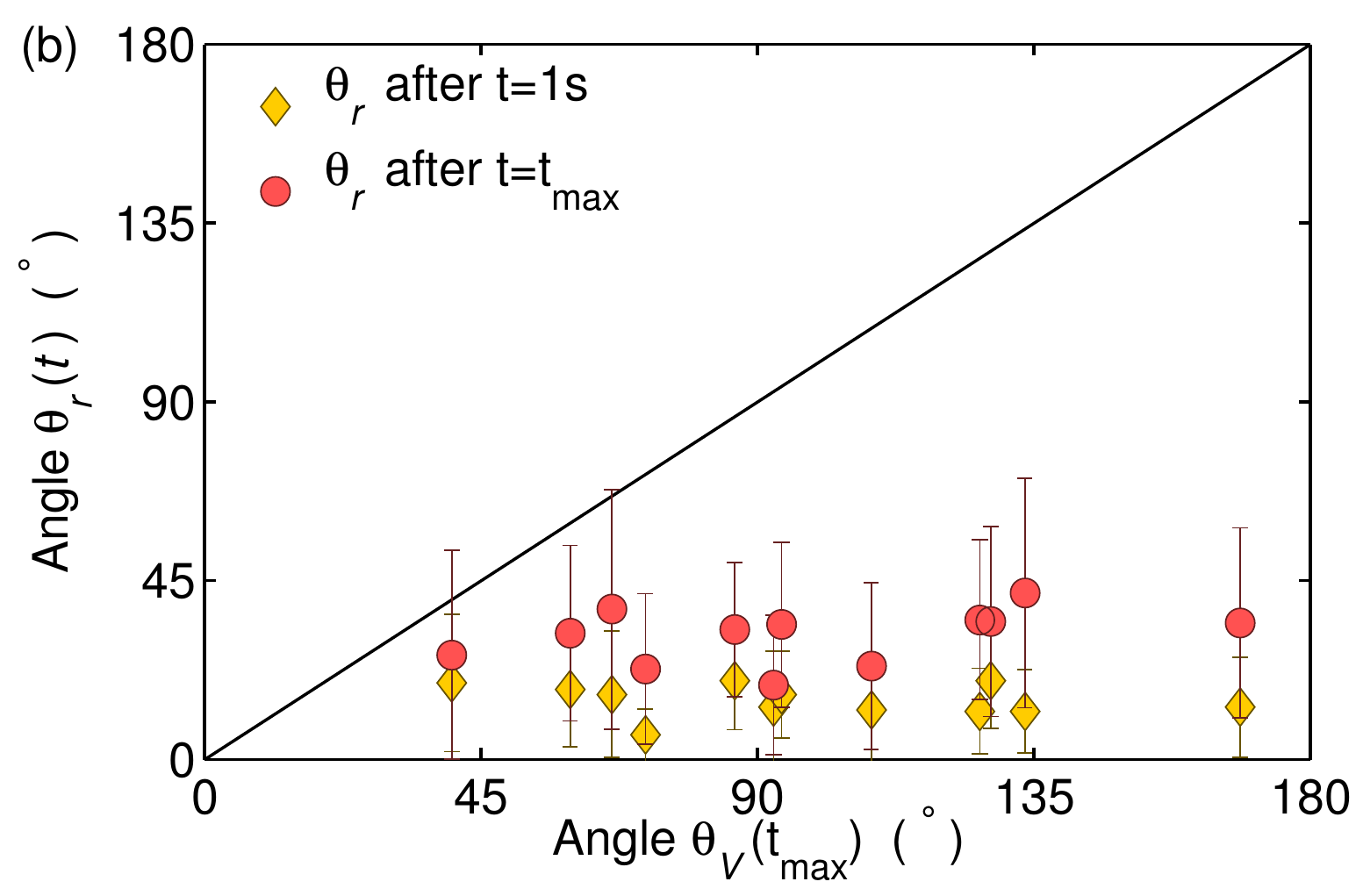}
 \caption{
{\bf Equal radius turning. }
(a) Temporal evolution of the angles $\theta_r(t)$, which measures the angular variation of the structural network of birds (dashed blue line), and $\theta_V(t)$, the angle between the flock's velocity ${\bf V}(t)$  and ${\bf V}(t=0)$ (solid red line), for turning event E1 (flock 20110208\_ACQ3). 
(b) For all 12 events, on the x-axis we plot the maximal change of the flock's travel direction $\theta_V(t_{\rm max})$ calculated at final time t=$t_{\rm max}$ of the acquisition, while on the y-axis we plot two values of the angular variation of the network $\theta_r$ calculated at: (i) $t=1$s after the start of the acquisition (yellow diamonds), and (ii) at the final time of the turning event $t=t_{\rm max}$ (red circles).
All turning events show similar behavior with the angular variation $\theta_r$ being small compared to the maximal change of the direction of motion $\theta_V$. 
}
\label{fig:equal-radius}
\end{figure}

Trajectories of individual birds of a flock performing a collective turn show that birds follow paths of very similar radii of curvature, causing their trajectories to cross (see Fig.\ref{fig:trajectories}). 
We can be more quantitative in our characterization of this phenomena. Let us call  ${\bf r}_{ij} = {\bf r}_{j}-{\bf r}_{i}$ the position of bird $j$ relative to bird $i$. We can monitor the orientational change of ${\bf r}_{ij}$ between the initial time $t=0$ of the event and a later time $t$. This is described by the angle $\theta_{ij}$ between the two vectors ${\bf r}_{ij}(0)$ and ${\bf r}_{ij}(t)$
\begin{equation}
\cos\theta_{ij}(t)=\frac{{\bf r}_{ij}(0)}{\|{\bf r}_{ij}(0)\|}\cdot\frac{{\bf r}_{ij}(t)}{\|{\bf r}_{ij}(t)\|}~.
\label{eqn:costhetaij}
\end{equation}
Averaging over all pairs of birds $i$ and $j$, we obtain $\theta_r=\langle\theta_{ij}\rangle_{i,j}$, which quantifies the angular variation of the whole structural network of birds over time.
 In Fig.~\ref{fig:equal-radius}-a we show the evolution over time of  $\theta_r(t)$ and we compare it with $\theta_V(t)$, the angle between the global direction of motion of the flock at time $t$ and the one at time $t=0$. The interesting point is that for parallel path turning the angles $\theta_r$ and $\theta_V$ must be highly correlated, with $\theta_r=\theta_V$ in the rigid-body limit. On the contrary, we find that while the angle $\theta_V$ increases strongly during the turn, measuring a change of the flock's direction of motion, angle $\theta_{r}$ does not change considerably. Other turning events show similar behaviour, with small changes in $\theta_r$ over the time scales of the analyzed turns, see Fig.\ref{fig:equal-radius}-b.
Small changes in the angular variation of the network of the flock mean that the birds adapt the new direction of motion with no changes in their relative positions with respect to the absolute reference frame, a characteristic of the equal radius path turning. We show this illustratively in Fig.\ref{fig:elongation}-a,b. 
As discussed in the main text of the paper, reorientation of the birds in the flock in this way not only enables efficient and fast turning of a flock as a collective, it also significantly changes the positional role of individuals within the flock, so as to alternate risky positions over time. 
  
\newpage
\section*{Individual deviations from the global direction}

\vspace{0.4 cm}
\noindent
{\bf Dealignment time factor}. We define the dealignment time factor $\delta_i(\tau)$ as a percentage of time during a time interval of length $\tau$ before the start of the turn $t_1$, during which bird $i$ deviates from the mean direction of motion more than a given threshold value $C_0$
\begin{equation}
 \delta_i(\tau) = \frac{1}{\tau}\sum_{t=t1-\tau}^{t_1}\Theta\left(C_0- C_i(t) \right).
 \label{eq:dealign-time-factor}
\end{equation}
Here, $\Theta(x)$ is a Heaviside  step function, so that $\Theta(x)=0$ for $x<0$, and $\Theta(x)=1$ for $x>0$.
For the threshold value $C_0$, we choose a median of the set of values $\{C_i(t) | i=1,...,N; t \in \left[t_1-\tau,t_1\right]\}$. However, other thresholds can be also considered. For the choice of a median value as a threshold $C_0$, the dealignment time factor $\delta_i(\tau)$ gives the percentage of time during the time interval $\tau$ for which bird $i$ was more dealigned with the global direction of motion than at least $50\%$ of other birds in the flock.

\begin{figure}[h] 
  \centering
\includegraphics[width=0.4\columnwidth]{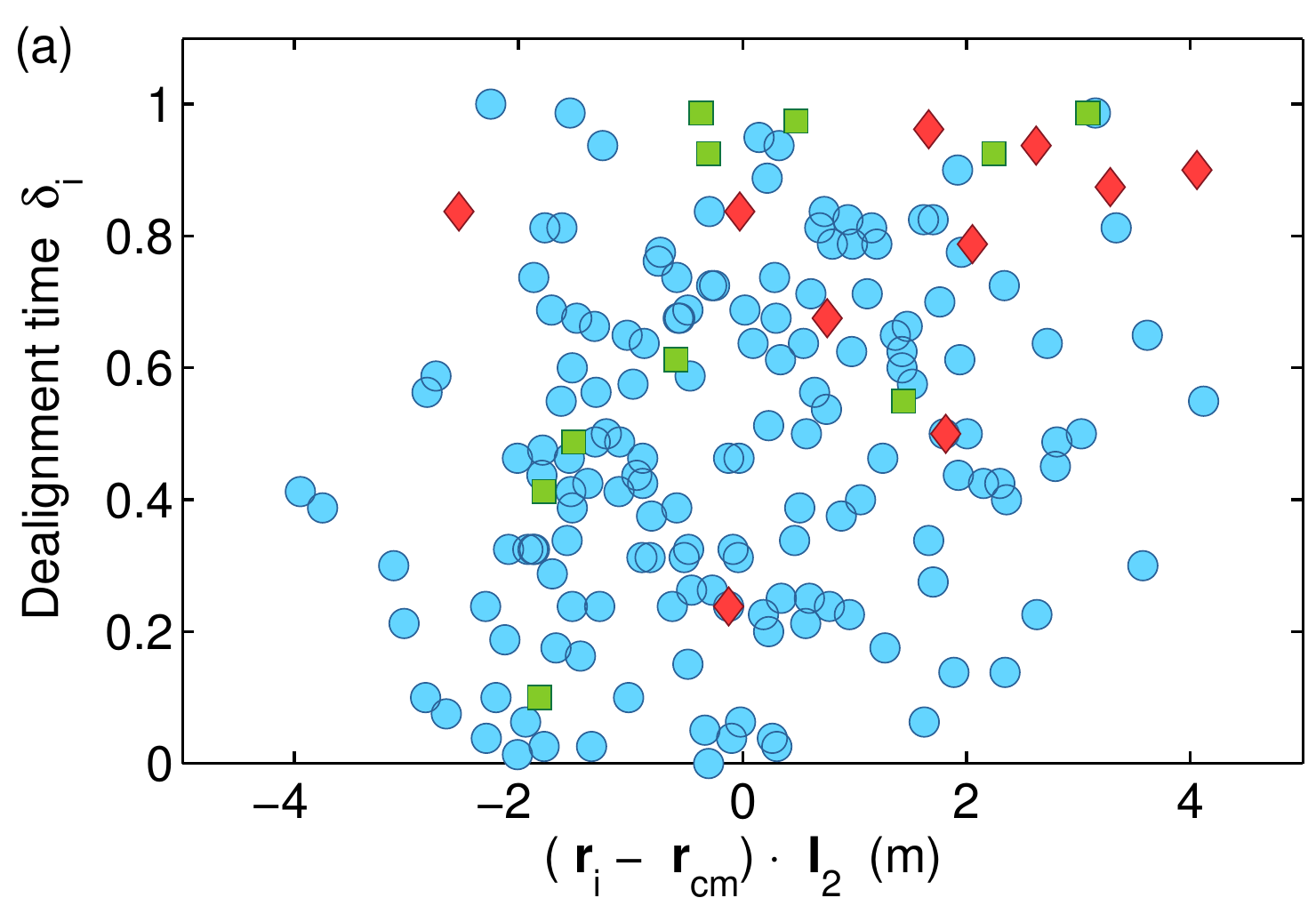}
\includegraphics[width=0.4\columnwidth]{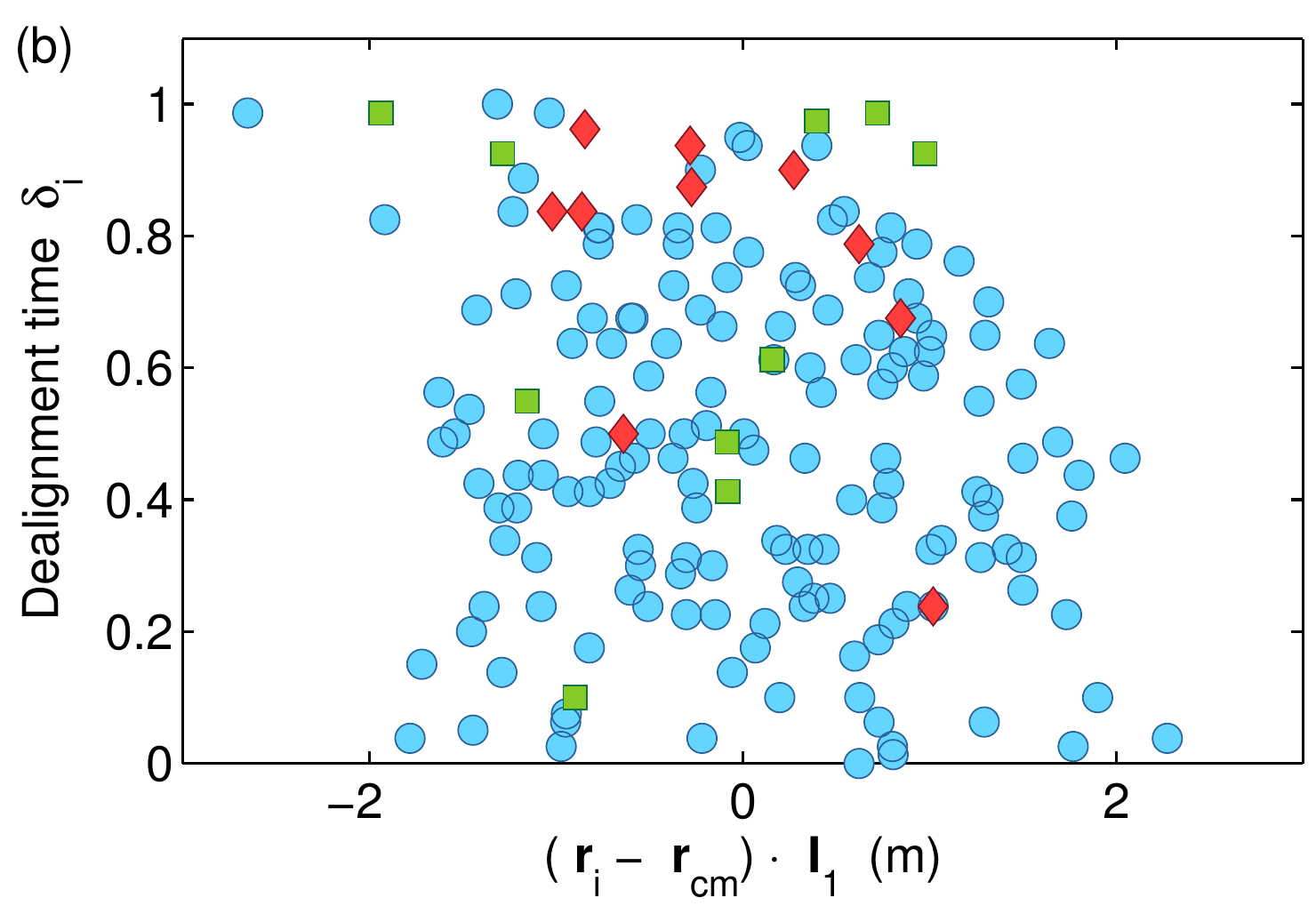}
\\
\includegraphics[width=0.4\columnwidth]{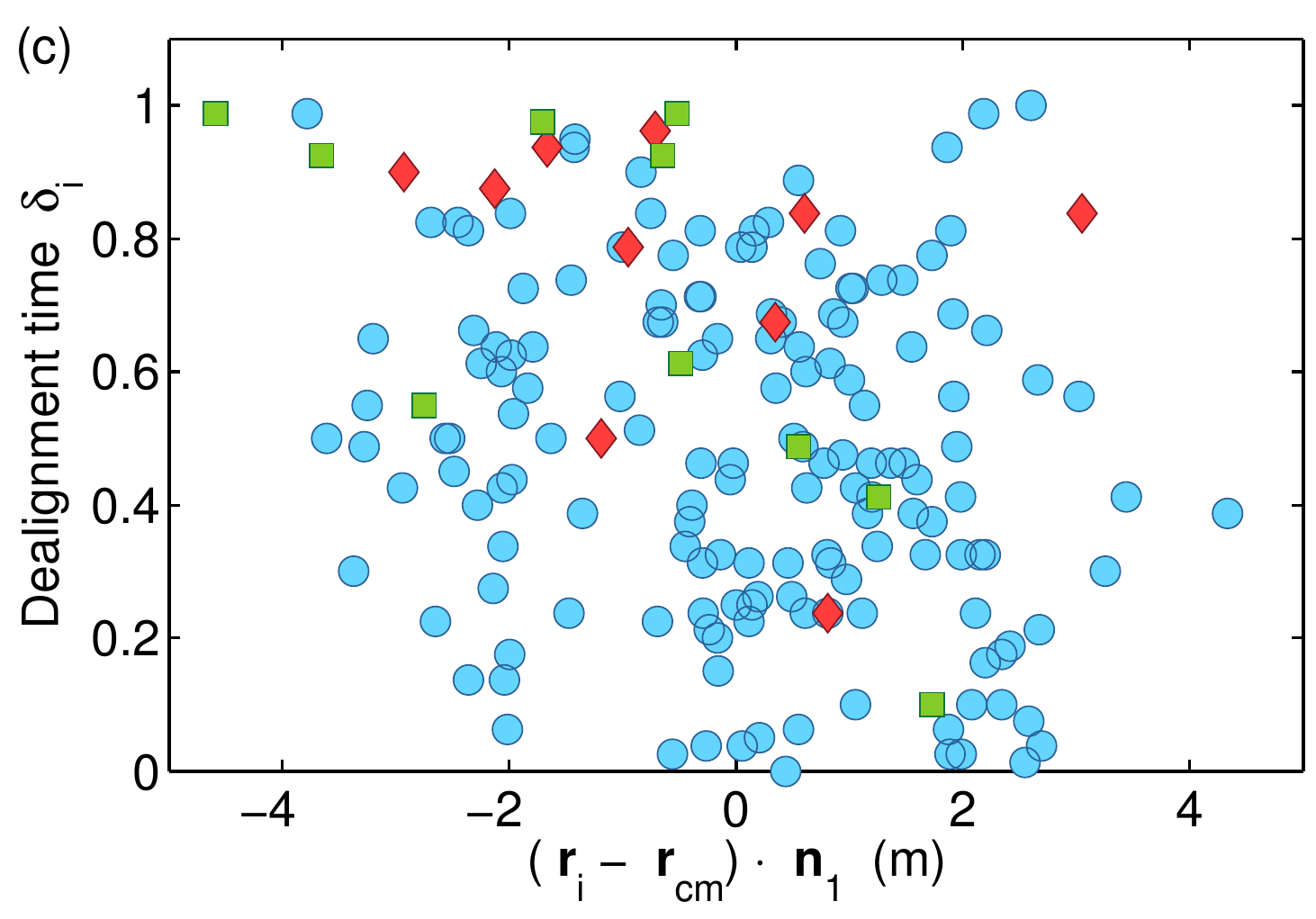}
\includegraphics[width=0.4\columnwidth]{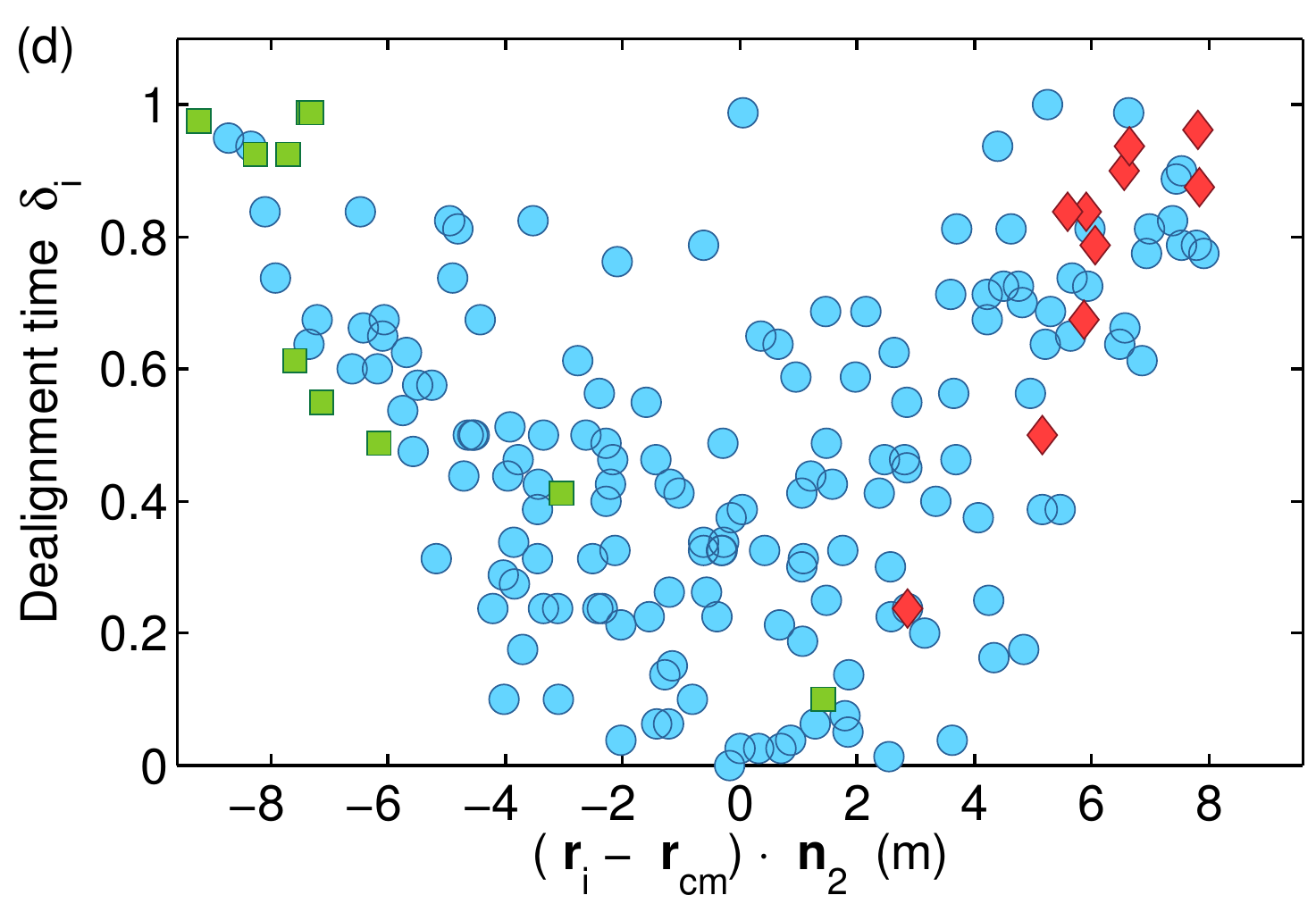}
 \caption{
{\bf Dependence of the bird's deviations from the global direction of motion on its location within the flock.} 
For event E1, dealignment time factor $\delta_i(\tau)$ is plotted as a function of position of bird $i$ along the elongation axes (a) ${\bf I}_2$ and (b) ${\bf I}_1$, with respect to the flock's center of mass at the moment the turn is initiated $t_1$. No strong dependence between the dealignment time factor $\delta_i(\tau)$ and position along these two elongation axes is found.
We also check the dependence on the position along (c) the direction of motion (front-back), and (d) transverse direction (left-right), the latter being almost the same as the longest elongation axis ${\bf I}_3$. 
The top ten ranked birds (red diamonds) and the last ten birds to turn (green squares) are marked.
 }
\label{fig:deviations-SI}
\end{figure}

As discussed in Section \ref{section:deviations}, we find a strong correlation between the location of an individual within the flock and the value of its dealignment factor: the farther away a bird is along the lateral elongation axis ${\bf I}_3$, the more frequently it exhibits  consistent fluctuations from the global direction of motion. We do not find, however, any obvious correlation when looking at the birds' positions along the other two elongation axes, as shown in Fig.~\ref{fig:deviations-SI}-a,b where we plot $\delta_i(\tau)$ as a function of the bird's position along other two axes ${\bf I}_2$ and
${\bf I}_1$. There is also no correlation between $\delta_i(\tau)$ and the birds' position along the direction of motion ${\bf n}_1$, that is, whether it is situated in the front or in the back of the flock (see Fig.~\ref{fig:deviations-SI}-c). The only correlation can be found between $\delta_i(\tau)$ and the lateral (left-right) direction ${\bf n}_2$ direction, which is not surprising as we showed that the longest elongation axis ${\bf I}_3$ is almost parallel to it at the start of the turn $t_1$,  (see Fig.~\ref{fig:deviations-SI}-d). 

\noindent
{\bf Dealignment amplitude}. We also define a normalized dealignment amplitude $\Delta C_i/C_0$, where $\Delta C_i$ is a maximal deviation from the mean flock's direction of motion during the time interval $\tau$ before the start of the turn
\begin{equation}
 \Delta C_i = C_0 - \min\{C_i(t) | C_i(t)<C_0; t \in \left[t_1-\tau,t_1\right]\}.
 \label{eq:dealign-amplitude}
\end{equation}	
We note that the velocities of individual birds ${\bf v}_i(t)$ are calculated from the individual trajectories of each bird $i$, by looking at its positions at the frequency of $10$hz and not at the sampling frequency $170$hz  (however, we used all the available data points sampled at $170$hz). This is because, as stated above, at such a high frequency of $170$hz the deviations from the mean direction of motion would include the zig-zag part of the trajectories which are due to wing-flapping. We avoid this by choosing to calculate the velocities at the time interval of $0.1$s, i.e.\ at $10$hz which is exactly the frequency of the flapping. 
Finally, we present the data for the time interval $\tau =1$s before the start of the turn at moment $t_1$. However, for the turning events in which the turn starts at moment $t_1<1$s from the start of the acquisition, the time interval $\tau$ starts from the beginning of the acquisition and, therefore, includes also time $t_1$. Typically, we use time period of $\tau=1$s which is short enough so that the results are not strongly influenced by this lack of data before the start of the turn, but long enough to give good statistics for dealignment.

\begin{figure}[h] 
  \centering
\includegraphics[width=.4 \columnwidth]{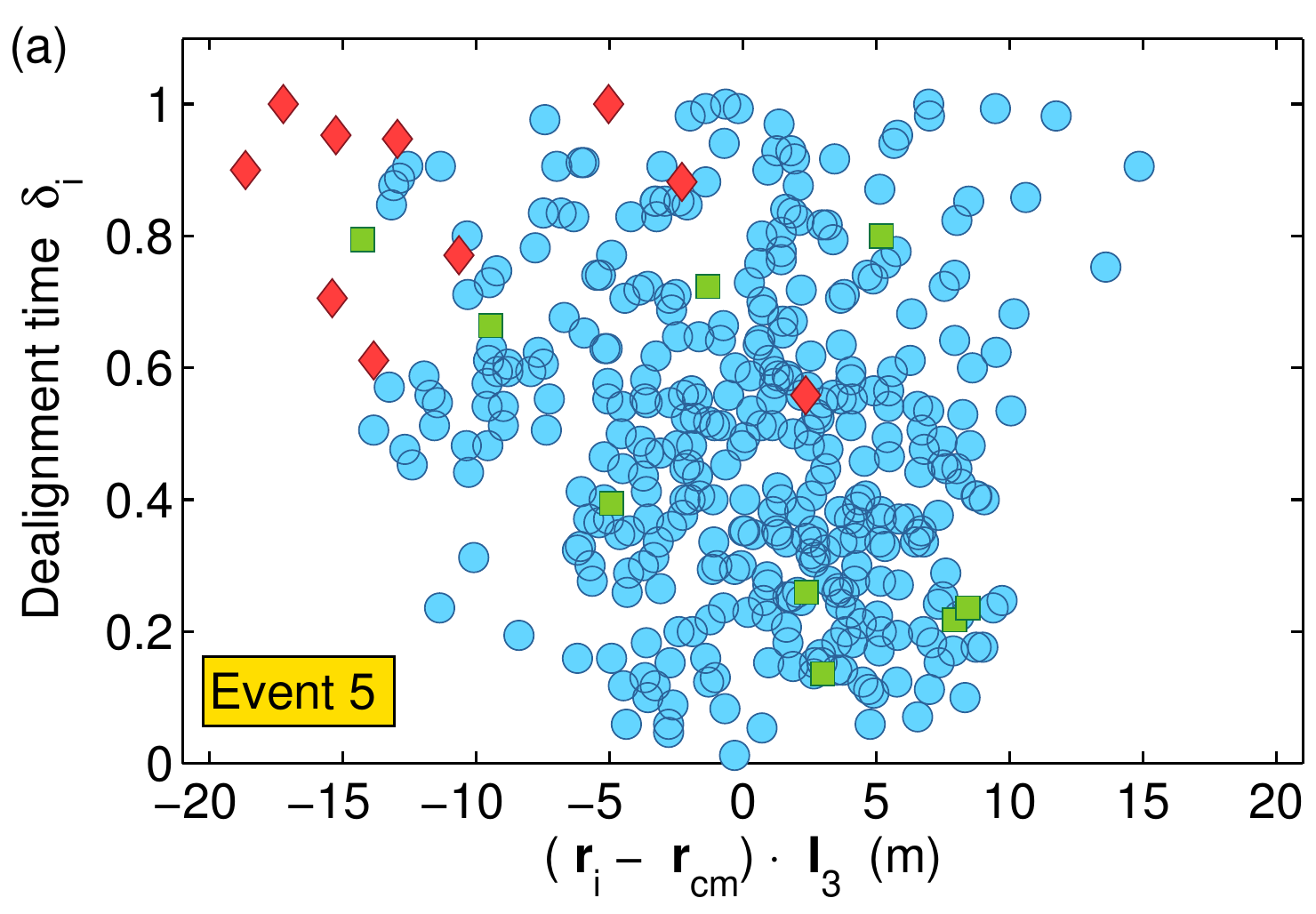}
\includegraphics[width=.4 \columnwidth]{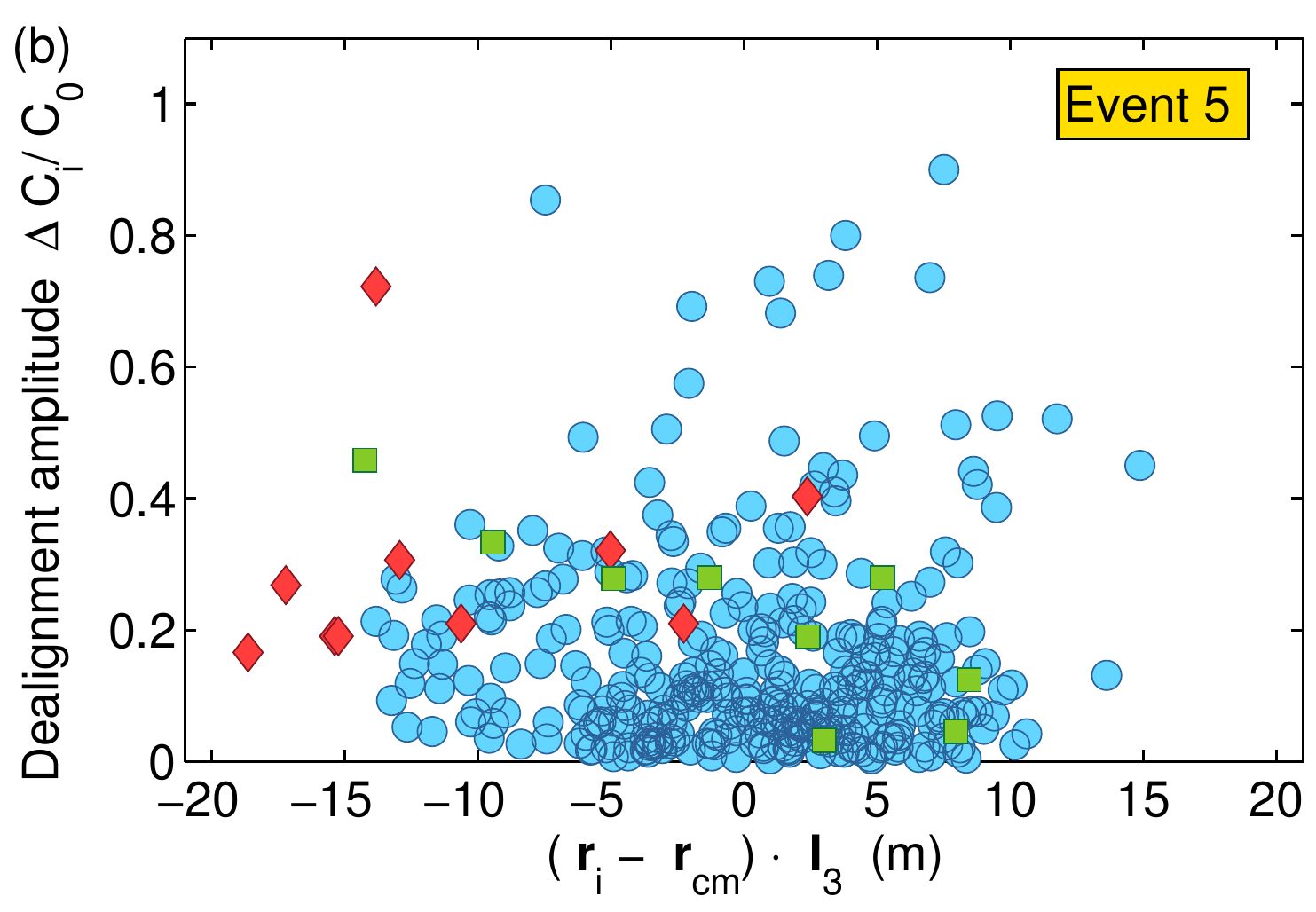}
\\
\includegraphics[width=.4 \columnwidth]{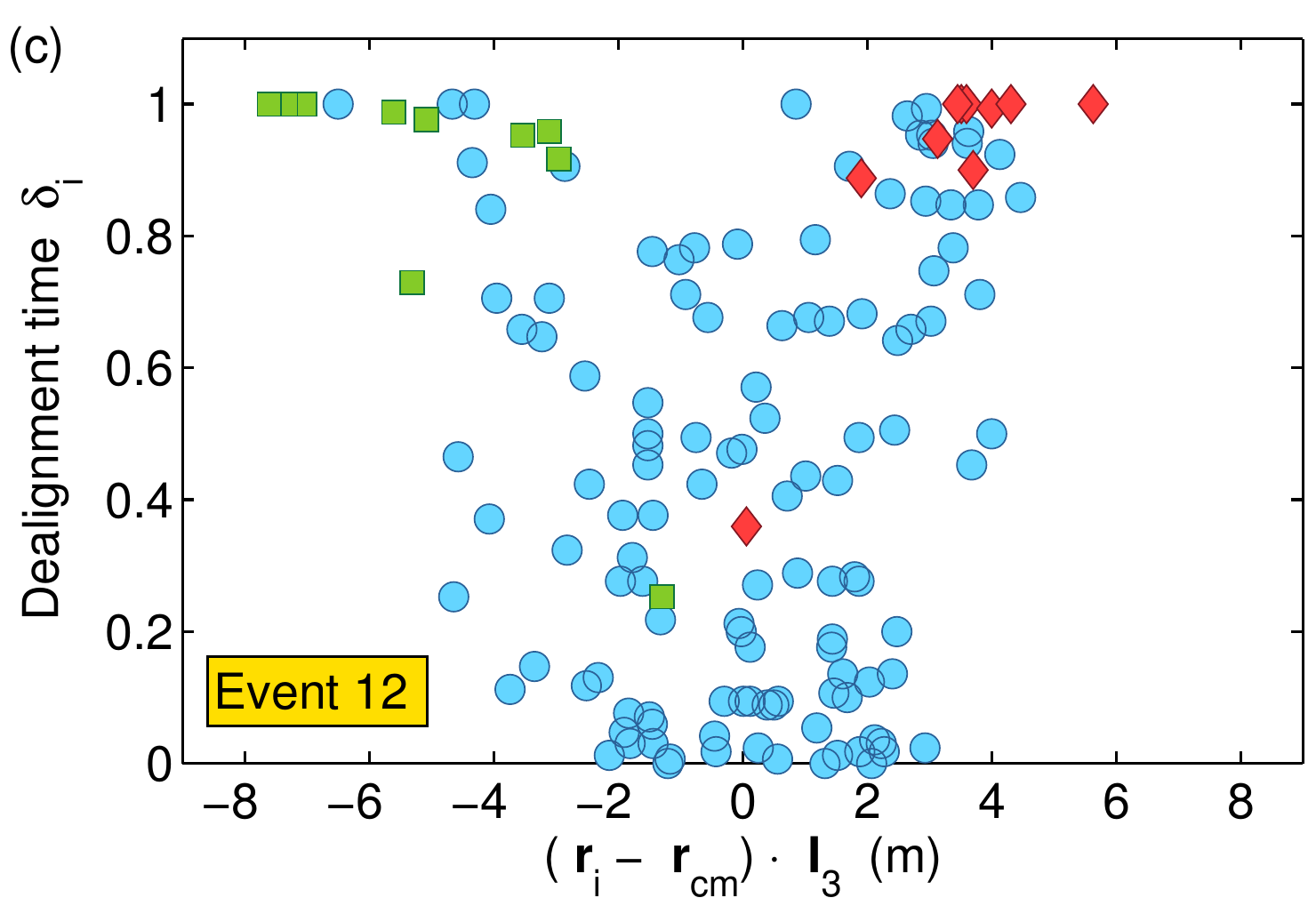}
\includegraphics[width=.4 \columnwidth]{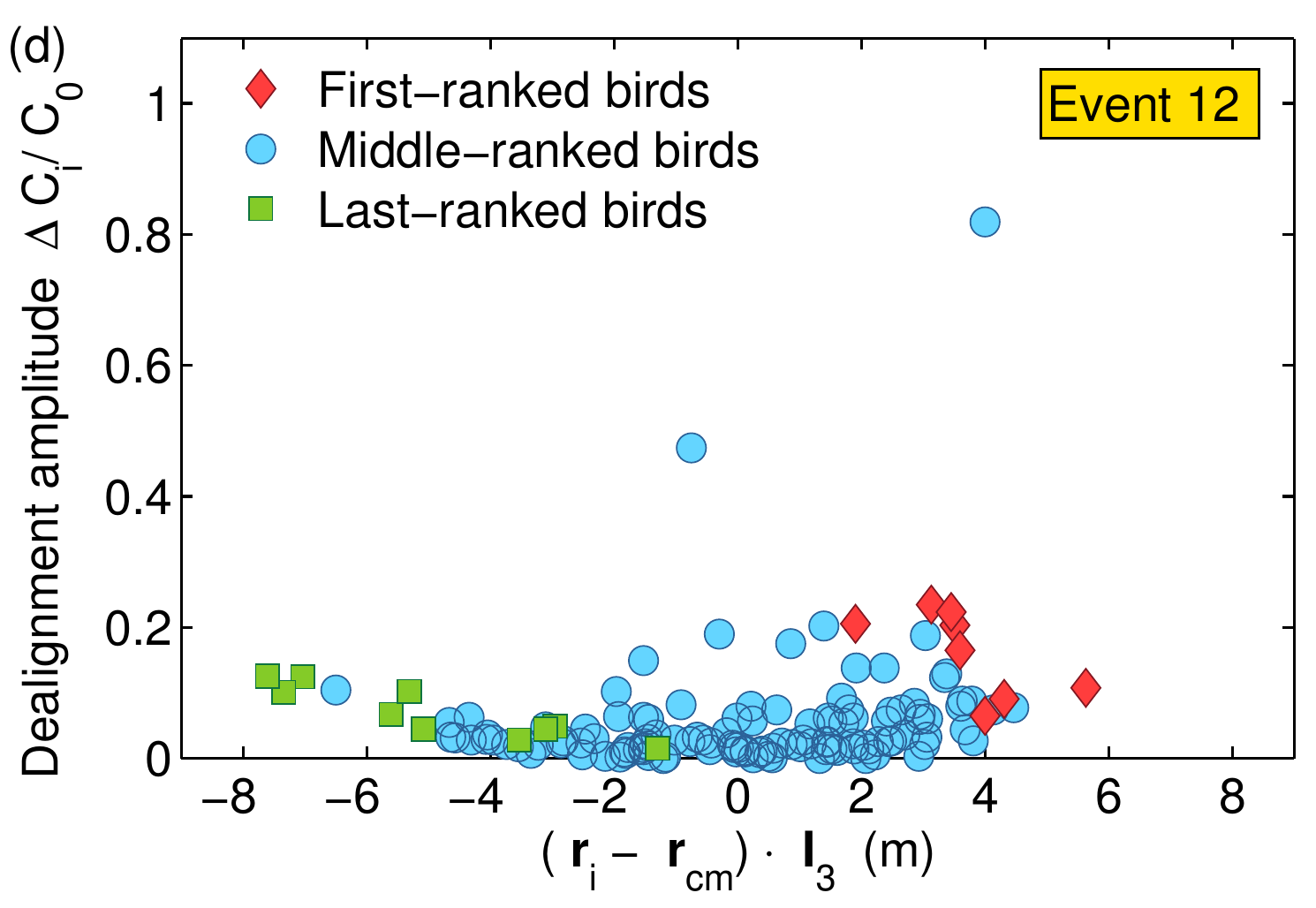}
 \caption{
{\bf Deviations from the global direction of motion for different turning events.} 
We show deviations from average direction of motion for two turning events E5 and E12.
(a), (c) Dealignment time factor $\delta_i(\tau)$ is plotted as a function of position of bird $i$ along the longest elongation axis ${\bf I}_3$, with respect to the flock's center of mass. On average, the farther away from the center of the flock a bird is, the longer time its directional correlation is lower than $C_0$.
The top ten ranked birds (red diamonds) are situated close to one edge, that is farthest away from the center of the flock along ${\bf I}_3$ axis, while last ten birds to turn (green squares) are close to the other edge of the flock or not, depending on a particular propagation of the turn through the flock.
(b), (d) Dealignment amplitude $\Delta C_i /C_0$ is shown as a function of position of bird $i$ along the longest elongation axis ${\bf I}_3$, with respect to the flock's center of mass. No obvious dependence on the position is observed.
 }
\label{fig:deviations-SI-different-flocks}
\end{figure}

In Section \ref{section:deviations}, we show that the top-ranked individuals, which are situated close to one edge of the flock, are  among the individuals with the highest realignment factor $\delta_i(\tau)$. On the contrary, we find that there is not such a correlation between the dealignment amplitude $\Delta C_i /C_0$ and the location of an individual within the flock. Here we show this for some of the other turning flocks, see Fig.~\ref{fig:deviations-SI-different-flocks}, confirming our conclusion that what triggers the turn is the presence of a repeated deviation from the flock's global direction and not the strength of a single, momentary deviation from the average behaviour.

\begin{figure}[h] 
  \centering
 \includegraphics[width=0.4\columnwidth]{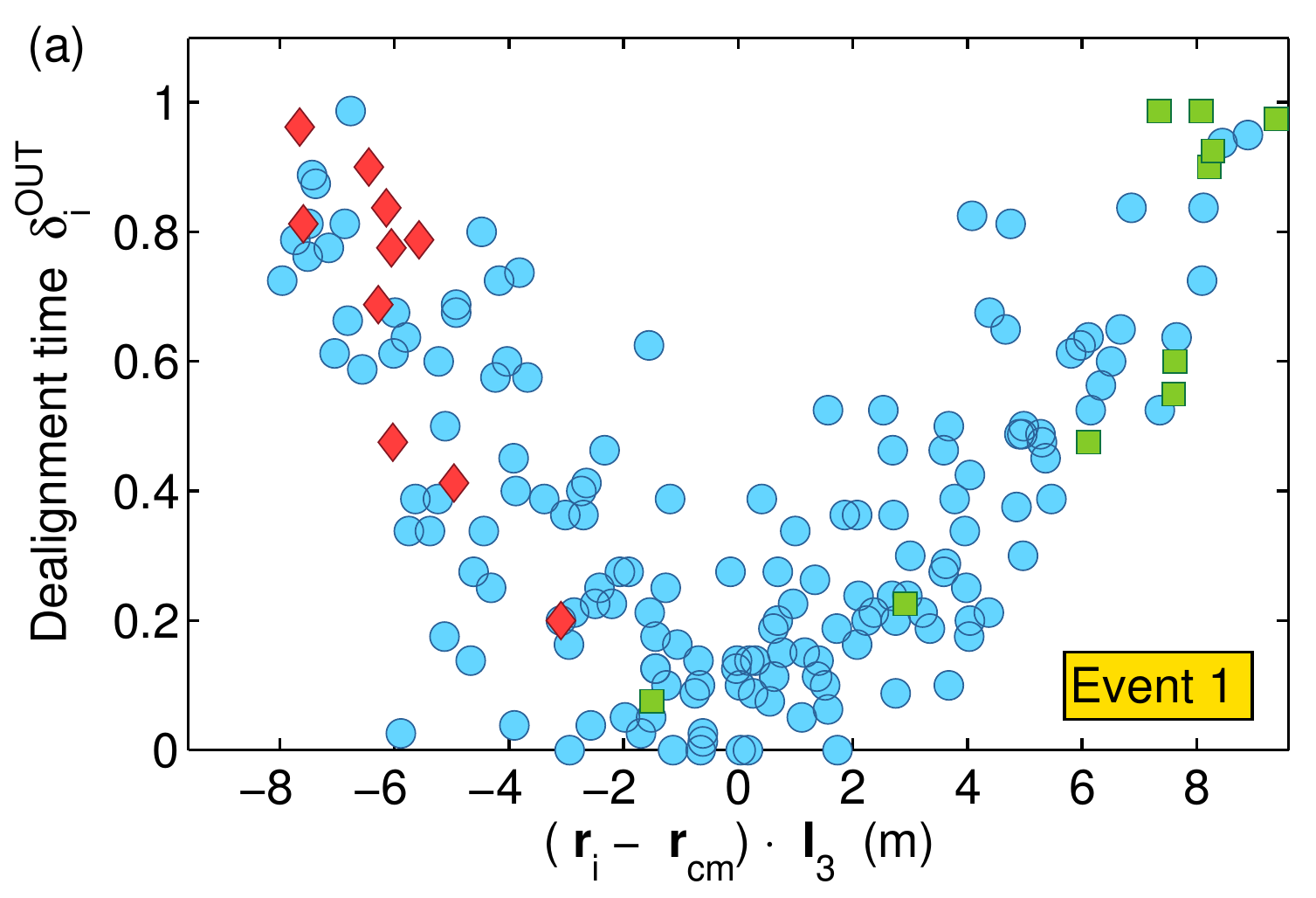}
\includegraphics[width=0.4\columnwidth]
{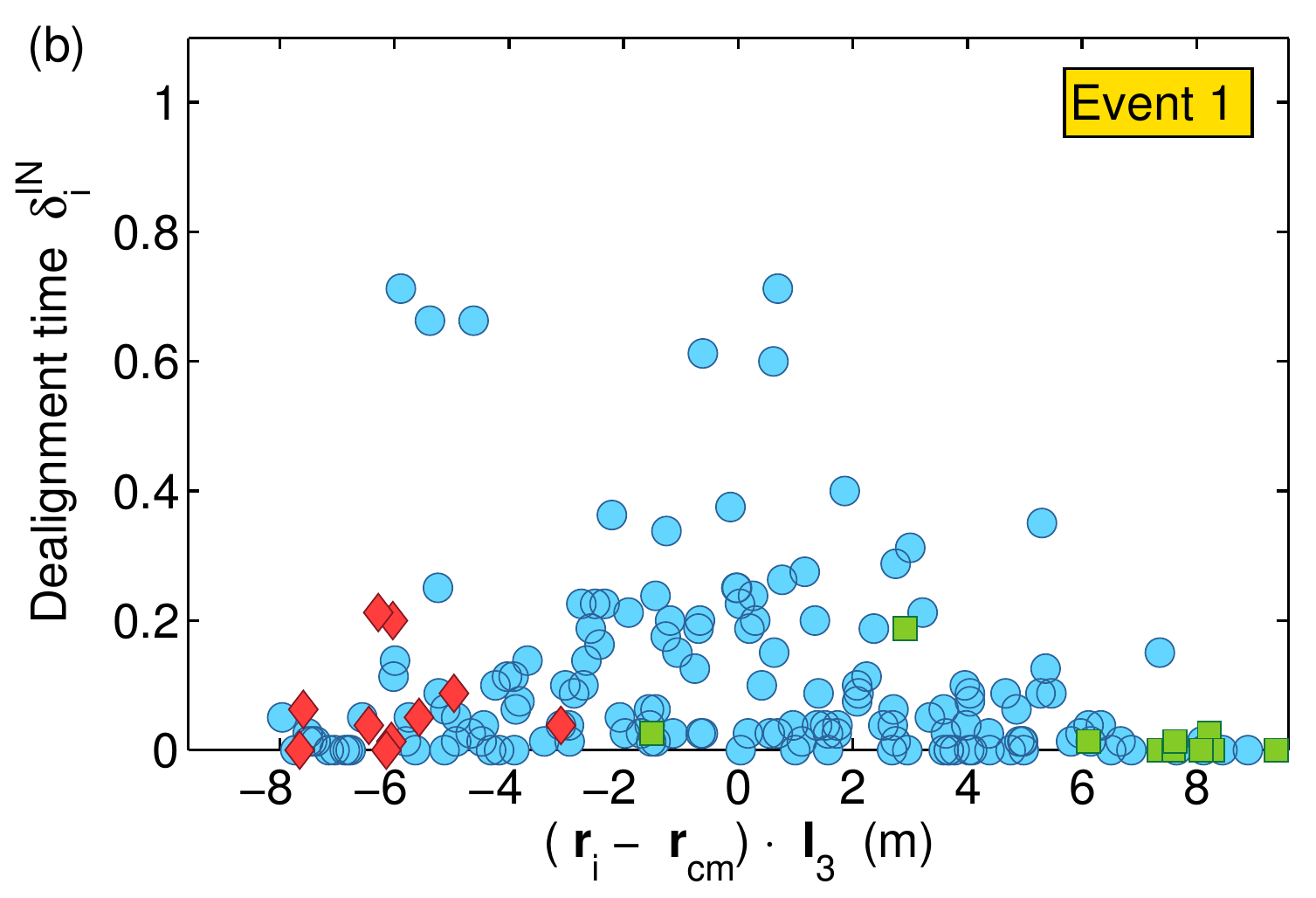}
\\
\includegraphics[width=0.4\columnwidth]{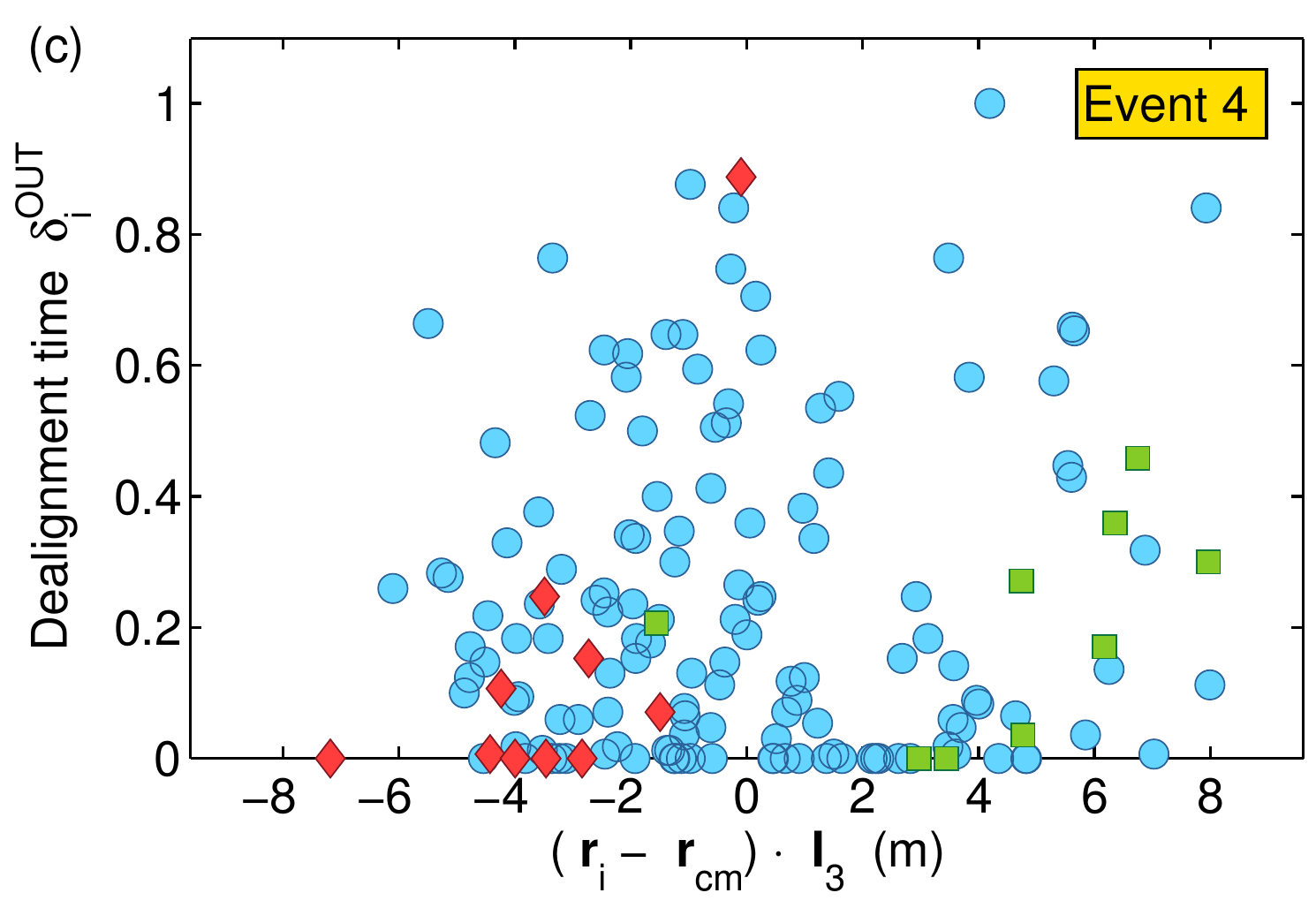}
\includegraphics[width=0.4\columnwidth]{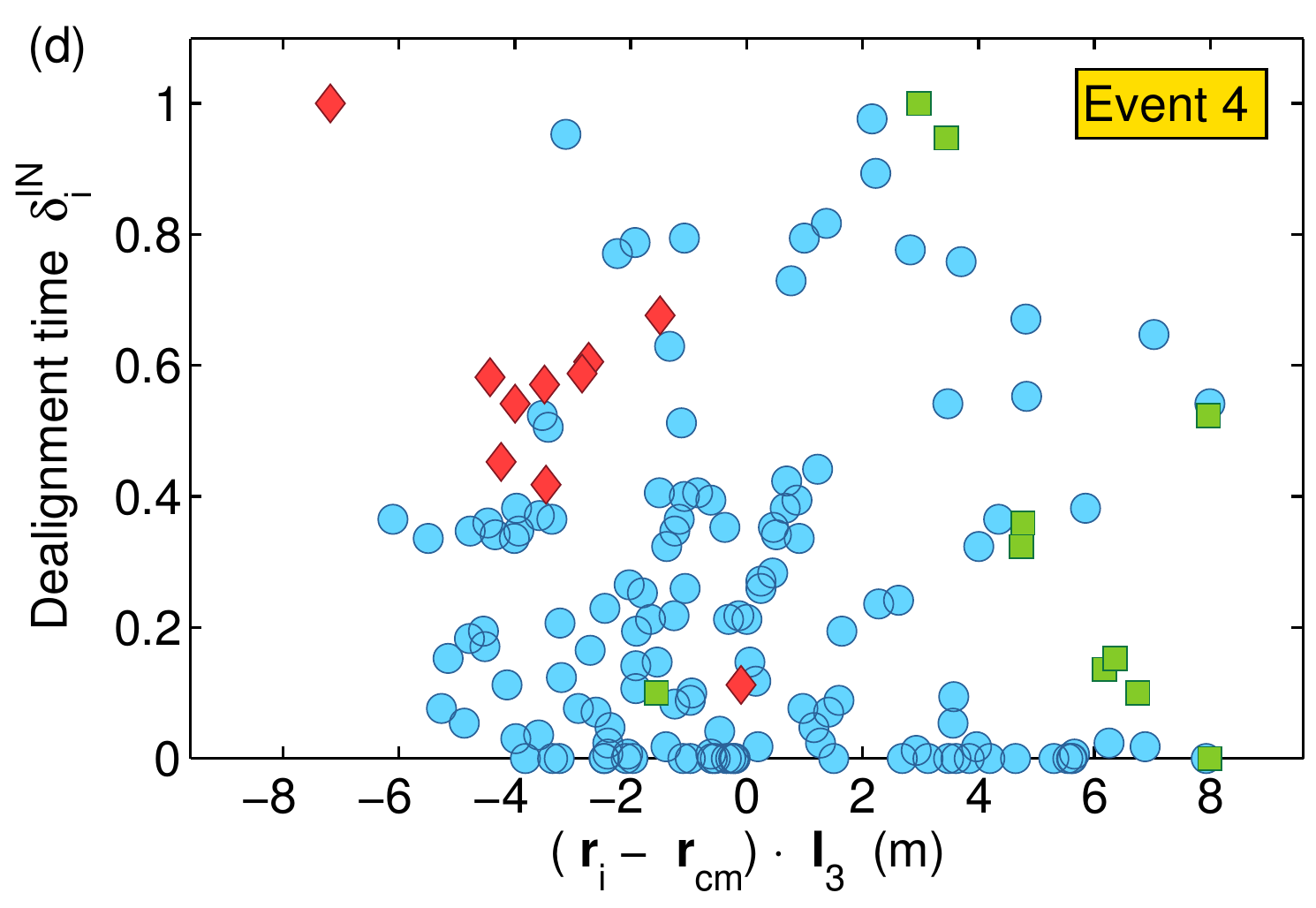}
\\
\includegraphics[width=0.4\columnwidth]{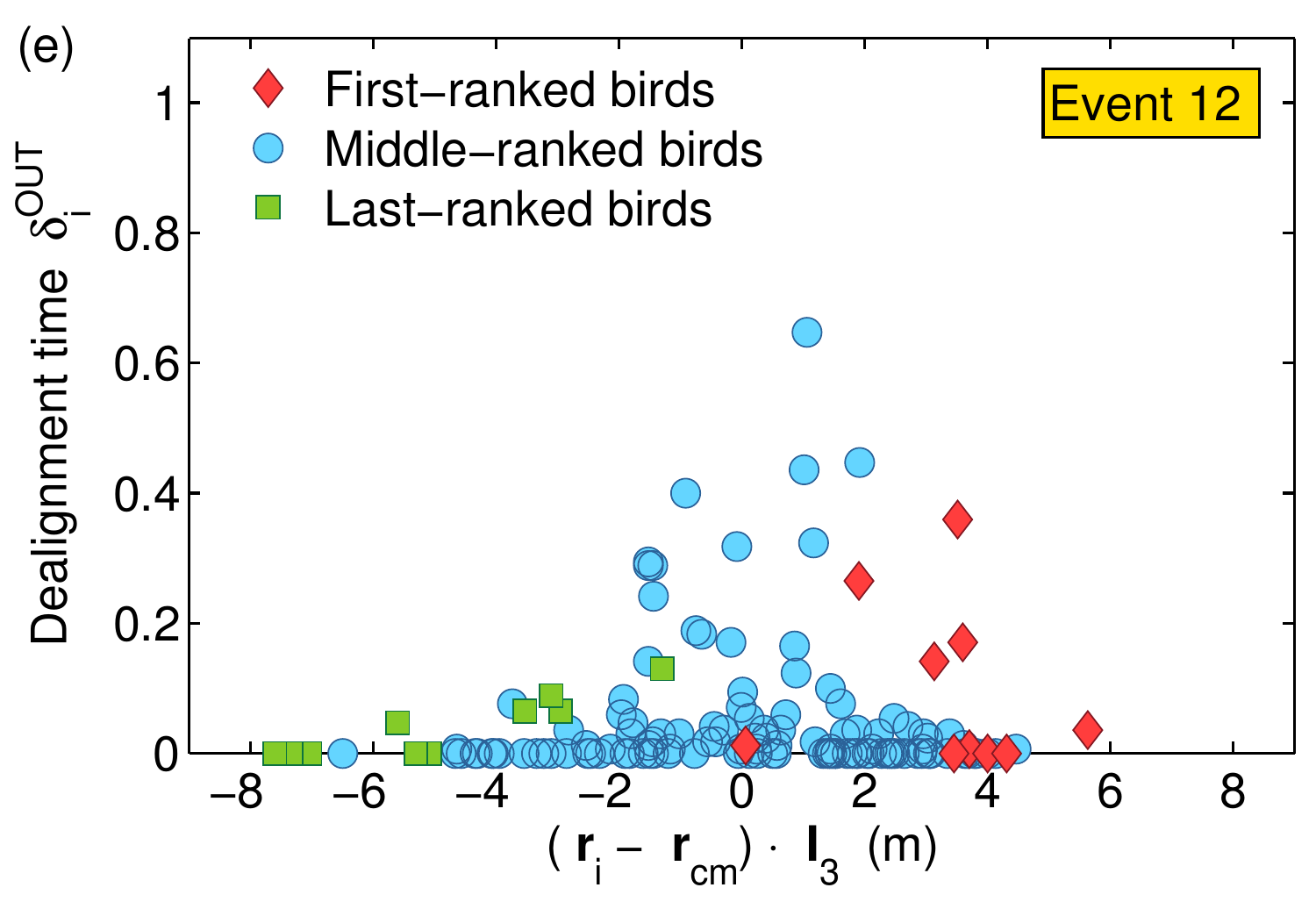}
\includegraphics[width=0.4\columnwidth]{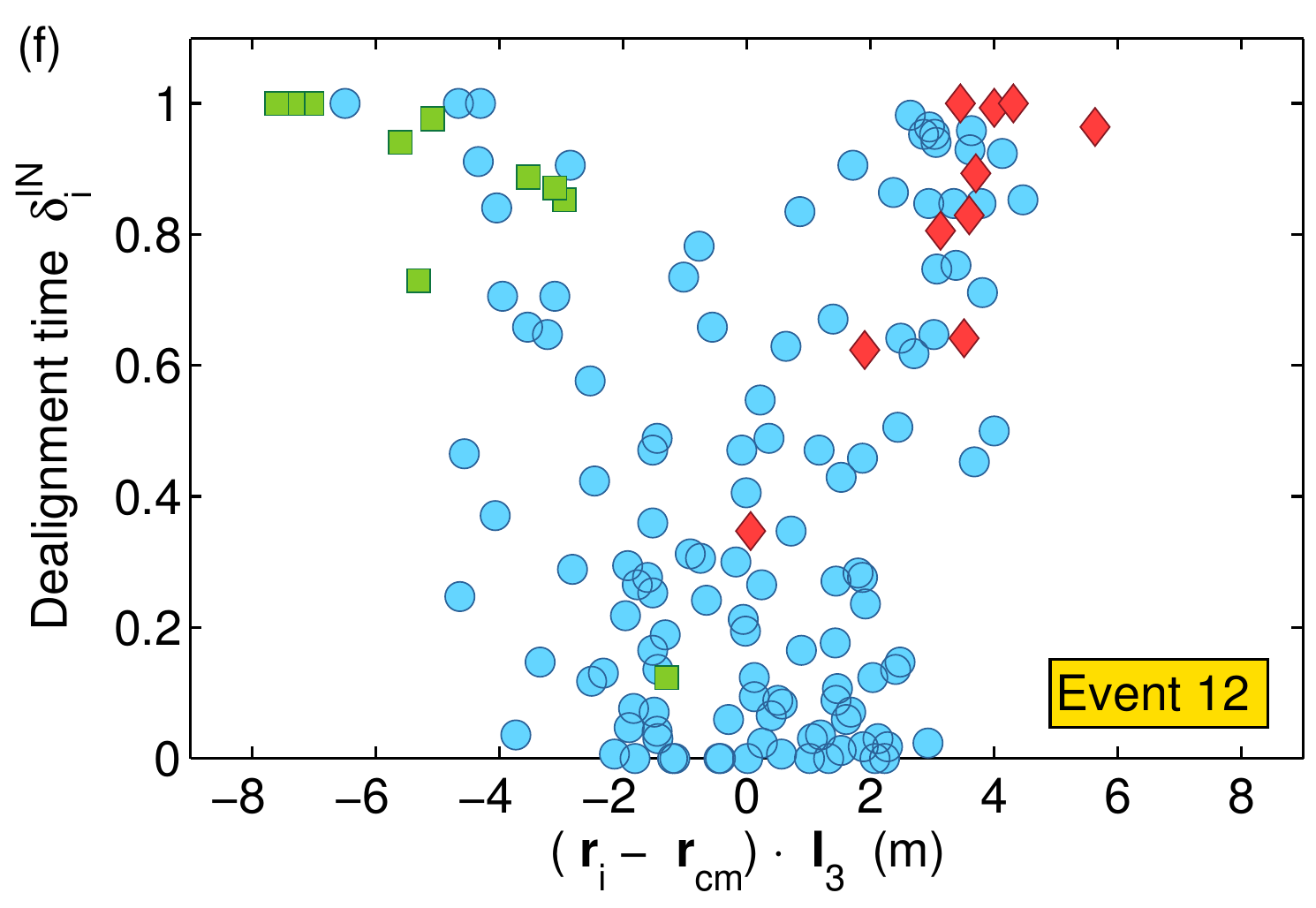}
 \caption{
{\bf Outward and inward deviations with respect to the center of a flock.} 
For each bird $i$ we separate deviations from the global direction of motion that are below the threshold value $C_0$ depending on whether they are directed outwards or inwards with respect to the center of the flock.
Appropriate dealignment time factors $\delta_i^{\rm out}$ and $\delta_i^{\rm in}$ are shown in panels (a)  and (b), respectively. The former one, $\delta_i^{\rm out}$, gives the percentage of time during time interval $\tau$ during which bird $i$ was deviating stronger than $C_0$ and in the direction away (outwards) from the flock. The latter one, $\delta_i^{\rm in}$, presents the amount of time within $\tau$ that bird $i$ was oriented more towards the center (inwards) of the flock.
At each moment of time $t$, we separate outward and inward deviations using a plane defined by the direction of motion ${\bf n}_1$ and the longest elongation axis ${\bf I}_3$. If within this plane we define a vector ${\bf n}_{\perp}$ orthogonal to the direction of motion ${\bf n}_1$, then the velocity deviations are calculated as ${\bf n}_{\perp}\cdot {\bf v}_i$. Depending on the position of the bird along the ${\bf I}_3$ axis with respect to the barycenter, and the deviation along ${\bf n}_{\perp}$, we can distinguish if a deviation is inwards or outwards.
For event E1, the results show that most deviations by the birds on the sides are made in the direction away from the flock and are responsible for the increase in the flock's elongation in the transverse direction. This can be compared to the diffusion close to the wall, where in the case of a flock the birds on the side have more space to deviate towards the sides (away from the center), than towards the center of the flock where most of their neighbors are. 
The turn starts by deviations in the opposite direction--towards the inside of the flock--by the two top ranked birds shown in panel (b) with the highest $\delta_i^{\rm in}$ among the top ranked birds (red diamonds).
On the other hand, birds that are positioned in the center of the flock, do not show any preference for deviations towards inside or outside of the flock, since they themselves are at the central part of the flock.
However, even though many of the events follow the trend as shown for event number E1, this is not always the case. For example, in event E4, there is no clear distinction between the outward and inward deviations. Finally, in event E12, we
find a situation opposite to event E1, where most of the deviations are inwards, and the turn starts by repeated outward fluctuations. 
}
\label{fig:deviations-IN-OUT}
\end{figure}

\newpage
\noindent
{\bf Outward and inward deviations}. 
In this paper we show that, while flying straight, flocks typically acquire asymmetric shapes, with their longest elongation axis being oriented perpendicular to the flock's velocity. Moreover, the results in Section \ref{section:deviations} demonstrate that birds deviate more and more frequently from the common flock direction, when being positioned closer and closer to the lateral edges of the flock. Here we contribute to these conclusions by distinguishing between the direction of these deviations, that is, whether they are directed towards the void space outside of the flock, or towards the inside of the flock where  the space is occupied with other birds. 
For many flocks we find that most of the deviations performed by the birds closer to the elongated edges are towards the boundary of the flock (outward), and not towards the center of the flock (inward), as for the turning event E1 shown in Fig.\ref{fig:deviations-IN-OUT}-a,b. 
This result is consistent with the maximal diffusion along the wing's axis found in \cite{cavagna+al_13a},
and, in particular, the dynamics of the border. 
As discussed therein, the dynamics of the edge birds results from balancing the availability of void space outside the flock, staying at the border of a flock rather than going astray \cite{lima_93}, and the reluctance of their internal neighbours to give up a more favourable position.
Finally, as this effect builds up in time, some of these birds successfully manages to move inwards, compensating the feedback to average motion,. This incites a response of their neighbours, followed by a collective turn of the whole flock.
In fact, in some of the turning events, we could clearly detect the inwards fluctuations performed by the top-ranked birds, which are the cause of the turn. In Fig.~\ref{fig:deviations-IN-OUT} we also show the distinction between outwards and inwards deviations from the common direction for two other events for which we noticed different pattern than the one discussed here. In particular, for event E12, the turn follows the joining of two separate flocks shortly before which could be the reason for it differing from the observed standard scenario.

\end{document}